\documentclass[conference]{IEEEtran}
\IEEEoverridecommandlockouts

\usepackage{tikz}
\usepackage{cite}
\usepackage{amsmath,amsfonts}

\usepackage{algorithmic}
\usepackage{graphicx}
\usepackage{textcomp}
\usepackage{xcolor}
\usepackage{color}
\usepackage{makecell}
\usepackage{multirow}
\usepackage{threeparttable}
\usepackage{booktabs}
\usepackage{tabularx}
\usepackage{xcolor}
\usepackage{url}
\usepackage{hyperref}
\usepackage{adjustbox}
\usepackage[most]{tcolorbox}

\usepackage{makecell}
\usepackage{multirow}
\usepackage{threeparttable}

\ifCLASSOPTIONcompsoc
\usepackage[caption=false, font=normalsize, labelfont=sf, textfont=sf]{subfig}
\else
\usepackage[caption=false, font=footnotesize]{subfig}
\fi

\newtcolorbox[%
auto counter]{mybox}[2][!h]{%
 enhanced jigsaw,
        colback=yellow!12,
 breakable,
 #1}

\definecolor{color1}{RGB}{255, 192, 122}
\definecolor{color2}{RGB}{232, 219, 210}
\definecolor{color3}{RGB}{219, 237, 251}
\definecolor{color4}{RGB}{255, 231, 214}

\begin{document}

\title{Asynchronous BFT Consensus Made Wireless

\thanks{
Corresponding author: Minghui~Xu (\href{mailto:mhxu@sdu.edu.cn}{mhxu@sdu.edu.cn})

This study was supported by the National Natural Science Foundation of China (No. 62302266, 62232010, U23A20302), the Shandong Science Fund for Excellent Young Scholars (No.2023HWYQ-008), and the project ZR2022ZD02 supported by Shandong Provincial Natural Science Foundation.}
}

\author{
	\IEEEauthorblockN{Shuo~Liu$^{\dag}$, Minghui~Xu$^{\dag}$, Tianyi~Sun$^{\dag}$, Xiuzhen~Cheng$^{\dag}$}
	\IEEEauthorblockA{$^\dag$ School of Computer Science and Technology, Shandong University}
}

\maketitle

\begin{abstract}
    Asynchronous Byzantine fault-tolerant (BFT) consensus protocols, known for their robustness in unpredictable environments without relying on timing assumptions, are becoming increasingly vital for wireless applications. While these protocols have proven effective in wired networks, their adaptation to wireless environments presents significant challenges. Asynchronous BFT consensus, characterized by its $N$ parallel consensus components (e.g., asynchronous Byzantine agreement, reliable broadcast), suffers from high message complexity, leading to network congestion and inefficiency, especially in resource-constrained wireless networks. Asynchronous Byzantine agreement (ABA) protocols, a foundational component of asynchronous BFT, require careful balancing of message complexity and cryptographic overhead to achieve efficient implementation in wireless settings. Additionally, the absence of dedicated testbeds for asynchronous wireless BFT consensus protocols hinders development and performance evaluation. To address these challenges, we propose a consensus batching protocol (ConsensusBatcher), which supports both vertical and horizontal batching of multiple parallel consensus components. We leverage ConsensusBatcher to adapt three asynchronous BFT consensus protocols (HoneyBadgerBFT, BEAT, and Dumbo) from wired networks to resource-constrained wireless networks. To evaluate the performance of ConsensusBatcher-enabled consensus protocols in wireless environments, we develop and open-source a testbed for deployment and performance assessment of these protocols. Using this testbed, we demonstrate that ConsensusBatcher-based consensus reduces latency by 48\% to 59\% and increases throughput by 48\% to 62\% compared to baseline consensus protocols.
\end{abstract}

\begin{IEEEkeywords}
Wireless consensus, Byzantine-fault tolerance, asynchronous, wireless networks.
\end{IEEEkeywords}

\section{Introduction}

Byzantine fault-tolerant (BFT) consensus is a critical component of applications that require agreement among a majority of network participants to resist Byzantine attacks. Asynchronous BFT consensus, emerging as a leading solution in recent wired networks, enhances robustness by removing reliance on timing assumptions. This characteristic renders asynchronous BFT consensus protocols highly advantageous in environments with unpredictable network delays, distinguishing them from synchronous and partially synchronous BFT consensus protocols. 
Wireless networks often experience varying signal strengths and interference, conditions under which the robustness of asynchronous BFT consensus protocols proves particularly beneficial.
For example, wireless applications that rely on reaching consensus as a prerequisite for initiating follow-up tasks include dynamic task allocation \cite{TaskAllocation}, collective map construction \cite{MapBuilding}, obstacle avoidance \cite{Obstacle}, and UAV/robot-assisted search and rescue \cite{rescue}.

In the domain of practical wireless BFT consensus protocols, numerous studies have focused on either synchronous wireless networks~\cite{reli, bhat2023eesmr} or partially synchronous wireless networks~\cite{TEE-1, TEE-2, 802.11_PBFT, BLOWN, wChain, bohm2024tinybft, liu2024partially}. Turquois~\cite{Turquois} can be applied to asynchronous wireless networks; however, it is a k-agreement protocol, meaning it does not guarantee termination for all honest nodes. LAP-BFT~\cite{kong2022lap} relies on a node detection mechanism, which carries the risk of being controlled by Byzantine nodes. Several well-established asynchronous BFT consensus protocols have emerged in wired networks in recent years, including HoneyBadgerBFT\cite{miller2016honey}, BEAT\cite{duan2018beat}, and Dumbo \cite{guo2020dumbo}. These protocols have demonstrated high performance and strong robustness in wired networks, marking a significant milestone for BFT consensus in asynchronous environments. This incentivizes us to explore the adaptation of these asynchronous BFT consensus protocols to wireless environments. In particular, we focus on the following challenges:


\textbf{[Challenge I]}
Asynchronous BFT consensus protocols involve $N$ parallel consensus components. In HoneyBadgerBFT and BEAT, there are $N$ parallel reliable broadcast (RBC)~\cite{bracha1987asynchronous, cachin2005asynchronous} instances and $N$ parallel asynchronous Byzantine agreement (ABA)~\cite{bracha1984asynchronous, cachin2000random} instances. In Dumbo, there are $N$ parallel provable reliable broadcast (PRBC)~\cite{guo2020dumbo} instances and $N$ parallel consistent broadcast (CBC)~\cite{cachin2001secure} instances.
While $N$ parallel consensus components result in high message complexity, this does not lead to network congestion in wired networks. In wired networks, any two nodes have stable channels and high bandwidth, allowing $N$ parallel consensus components to be efficiently managed by allocating $N$ separate threads. 
However, wireless networks do not have these advantages. Nodes in wireless networks share channels and need to compete for channel access to broadcast messages. When the consensus component protocol requires a node to send one message, it needs to compete for channel access once. With $N$ parallel components, the node must compete for channel access $N$ times. Therefore, running $N$ parallel consensus components amplifies channel contention by a factor of $N$, resulting in significant network congestion and reduced consensus efficiency. 

\textbf{[Challenge II]}
As a critical consensus component of asynchronous BFT consensus, ABA can be implemented using either a local coin~\cite{bracha1984asynchronous} or a shared coin~\cite{cachin2000random}, each presenting different trade-offs between message complexity and cryptographic overhead. Local coin-based ABA has higher message complexity, $\mathcal{O}(N^3)$, but avoids using computation-intensive cryptographic tools. In contrast, shared coin-based ABA has lower message complexity, $\mathcal{O}(N^2)$, but relies on computation-heavy threshold cryptography. Balancing these trade-offs in resource-constrained environments is not merely a selection but demands a tailored design approach to meet the unique requirements of asynchronous BFT consensus. This challenge is further amplified in asynchronous BFT consensus, which involves $N$ parallel ABA instances~\cite{miller2016honey, duan2018beat} or $N$ serial ABA instances~\cite{guo2020dumbo}, necessitating the development of efficient ABA implementations optimized for these configurations. Additionally, shared coin-based ABA cannot be directly deployed on resource-constrained devices without lightweight implementations of threshold cryptography. Current implementations, such as the Python version \cite{miller2016honey, duan2018beat, guo2020dumbo} and the C++ version\footnote{https://github.com/skalenetwork/libBLS}, rely on heavy dependencies and have significant storage overhead.


\textbf{[Challenge III]} 
There is a lack of practical testbed for evaluating asynchronous BFT consensus in wireless networks. While network simulators such as NS-3\footnote{https://www.nsnam.org/} and OMNeT++\footnote{https://github.com/omnetpp/omnetpp}, as well as practical testbeds like Flocklab2~\cite{FlockLab2} and FIT IoT-LAB~\cite{FIT-IoT-LAB}, are available, they lack high-level network APIs and consensus component APIs tailored for adapting asynchronous BFT consensus protocols from wired to wireless networks. High-level network APIs are essential for abstracting the complexities of underlying wireless networks, simplifying the adaptation of consensus protocols from wired to wireless networks. Consensus component APIs provide implementations of fundamental consensus components adapted for wireless networks, enabling developers to focus on designing and optimizing higher-level asynchronous BFT consensus. The absence of these interfaces increases development complexity and imposes higher time costs on developers, as they must address low-level implementation challenges instead of concentrating on core protocol design.

To tackle the aforementioned challenges, we propose ConsensusBatcher, a novel consensus batching protocol. ConsensusBatcher facilitates both vertical and horizontal batching of parallel and serial consensus components. Leveraging ConsensusBatcher, we construct a practical testbed to migrate three prominent BFT consensus protocols to wireless networks and assess their performance under various network conditions. Our contributions are highlighted as follows:

\begin{itemize}
    \item \textbf{Improving Consensus Through Batching.} 
    We propose ConsensusBatcher, a wireless communication protocol designed to manage multiple parallel instances in wireless networks efficiently. Unlike the direct deployment of parallel instances in wireless networks, ConsensusBatcher mitigates channel access competition and alleviates network congestion. Specifically, it optimizes the execution of $N$ parallel consensus components, a critical aspect of asynchronous BFT consensus. 

    \item \textbf{Lightweight ABA.}
    We provide lightweight ABA implementations, a critical component of asynchronous BFT consensus, optimized for resource-constrained environments. By developing efficient threshold cryptographic tools with lightweight libraries, we reduce computation and storage overhead, enabling the deployment of shared coin-based ABA in resource-constrained environments. Additionally, we optimize ABA for both $N$ parallel and $N$ serial instances, balancing message complexity and cryptographic overhead to meet the requirements of asynchronous BFT consensus.
    
    \item \textbf{Practical Testbed.}
    We present the design and implementation of a testbed for asynchronous wireless BFT consensus protocols. The testbed supports deploying five distinct consensus protocols, each based on different implementations of consensus components. By open-sourcing this testbed, we provide a valuable tool for both theoretical exploration and practical validation, fostering collaboration and further research in the field of asynchronous wireless consensus. Our testbed is open-source at \url{https://github.com/BDS-SDU/WirelessConsensus-Async}.

\end{itemize}

\section{Related work}

\textbf{Asynchronous BFT Consensus.} Asynchronous BFT consensus in wired networks has seen significant advancements in recent years, with a series of outstanding protocols emerging. HoneyBadgerBFT~\cite{miller2016honey}, a pioneering work, introduces a practical implementation using threshold encryption and Asynchronous Common Subset (ACS), which consists of $N$ parallel RBC instances and $N$ parallel ABA instances. BEAT~\cite{duan2018beat} built upon HoneyBadgerBFT by developing consensus protocols tailored to different scenarios through efficient component substitution. Dumbo~\cite{guo2020dumbo} addressed the high latency challenge of $N$ parallel ABA instances by utilizing multi-value validated Byzantine agreement (MVBA) to construct ACS. Subsequent research has extended these foundational works with components such as reproposable ABA~\cite{zhang2022pace} and Efficient MVBA~\cite{guo2022speeding,lu2020dumbo}. Existing asynchronous wireless BFT consensus protocols are not as well-developed as their wired counterparts. Turquois~\cite{Turquois} achieves k-consensus and can withstand dynamic message omissions. LAP-BFT~\cite{kong2022lap}, with its introduction of the node trusted status detection mechanism, achieves lightweight BFT consensus.


\begin{figure*}[!htb] 
\centering
  \subfloat[RBC \& PRBC (The blue lines represent the additional phase added to RBC in PRBC.)]{  \includegraphics[width=0.235\linewidth]{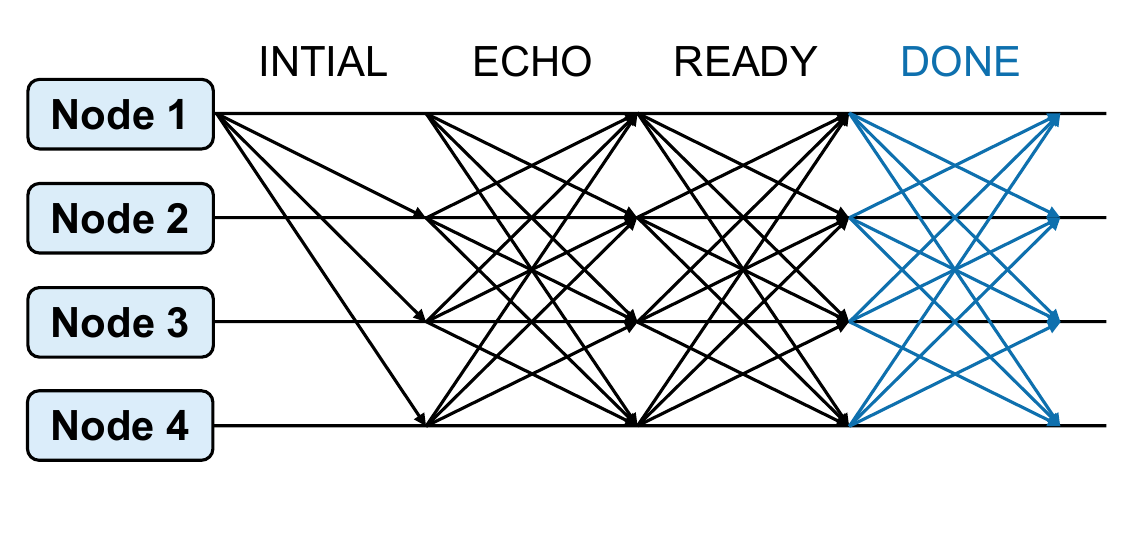}
\label{fig:RBC}}\hfill
  \subfloat[CBC]{  \includegraphics[width=0.235\linewidth]{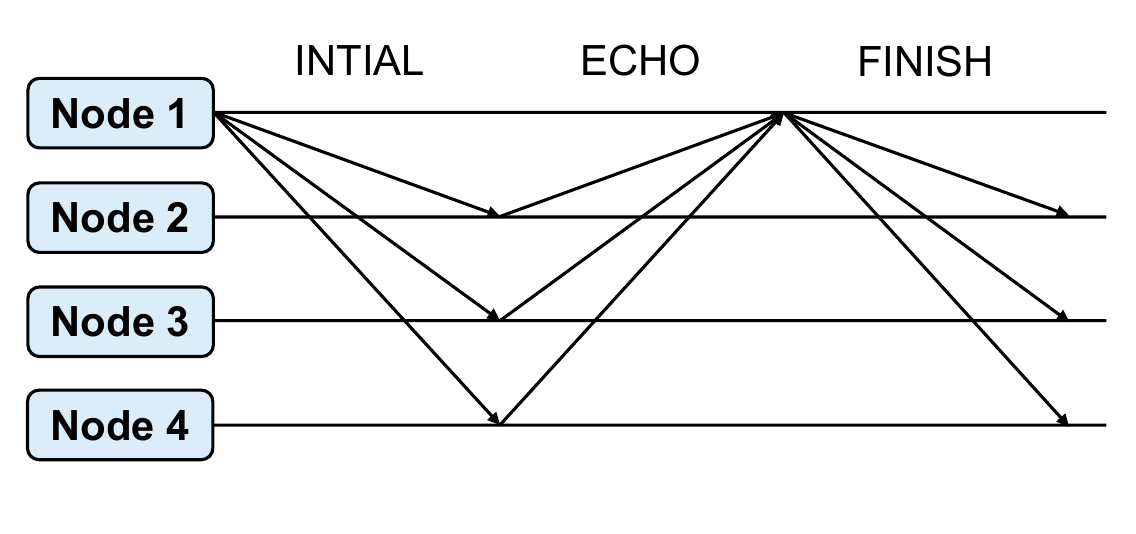}
\label{fig:CBC}}\hfill
    \subfloat[Round $r$ of Bracha's ABA]{  \includegraphics[width=0.235\linewidth]{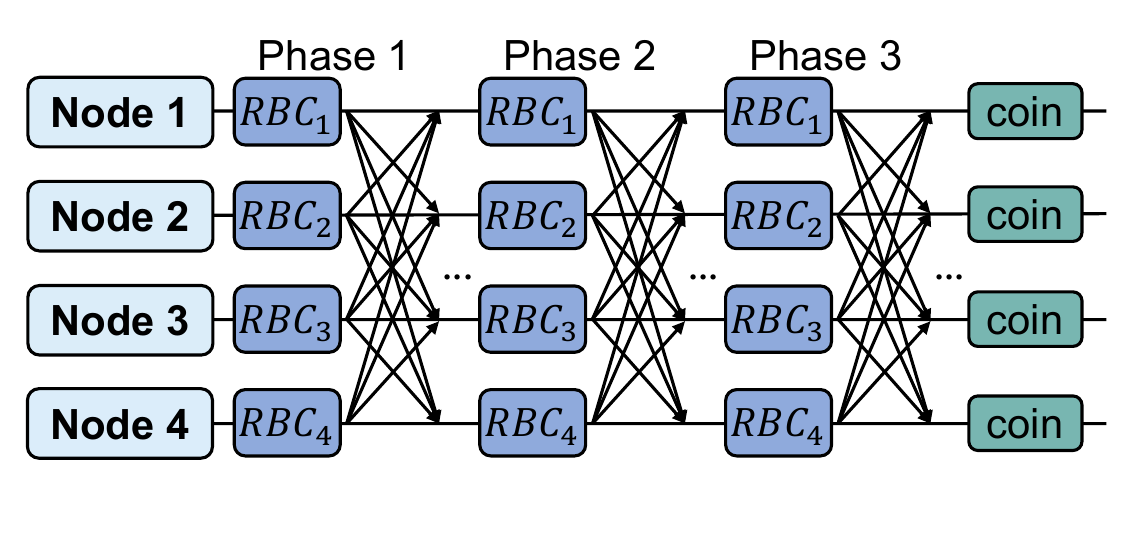}
\label{fig:Brachas-ABA}}\hfill
  \subfloat[Round $r$ of Cachin's ABA]{ \includegraphics[width=0.235\linewidth]{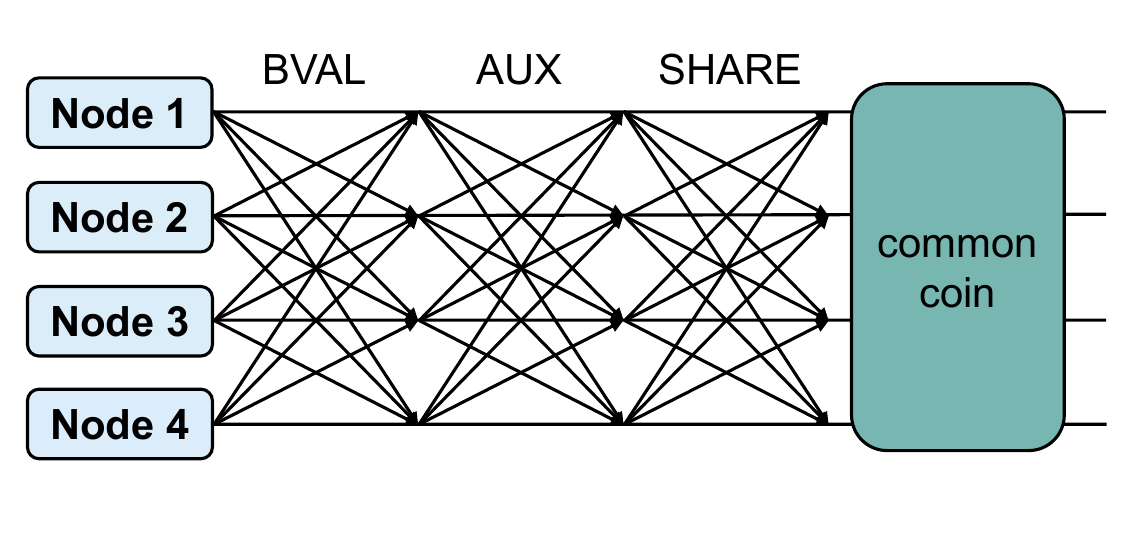}
\label{fig:Cachin'sABA}}\hfill
  \caption{Diagrams of consensus components}
  \label{fig:ABA_consensus} 
\end{figure*}

\textbf{(Partially) Synchronous Wireless BFT Consensus.} 
Synchronous wireless Byzantine Fault Tolerance (BFT) consensus protocols leverage synchronous communication mechanisms in their design. Goyal \textit{et al.} \cite{reli} introduce the ReLI framework, which facilitates Byzantine fault tolerance in resource-constrained IoT environments. Similarly, Bhat \textit{et al.} \cite{bhat2023eesmr} focus on enhancing energy efficiency within BFT State Machine Replication (SMR) protocols, targeting battery-powered devices in wireless ad hoc networks.

In contrast, partially synchronous wireless BFT consensus protocols address the dynamic and unpredictable nature of wireless networks. Trusted Execution Environments (TEEs) are utilized in several works to mitigate Byzantine faults in embedded wireless systems~\cite{TEE-1,TEE-2}. Zhou \textit{et al.} \cite{802.11_PBFT} evaluate the performance of wireless PBFT consensus protocols deployed over IEEE 802.11 networks.
The integration of wireless communication characteristics has led to novel protocol designs. For instance, the Proof-of-Channel consensus protocol, introduced in \cite{BLOWN}, exploits carrier sensing for single-hop ad hoc wireless networks. Additionally, a multi-hop ad hoc wireless consensus protocol, constructed around a spanner tree architecture, is proposed in \cite{wChain}.
Böhm \textit{et al.} \cite{bohm2024tinybft} address the implementation of BFT consensus protocols in environments with stringent memory constraints. Furthermore, Liu \textit{et al.} \cite{liu2024partially} propose the ReduceCatch protocol, enabling partially synchronous consensus protocols to transition from wired to wireless network contexts.

\section{Models and Preliminaries}
\subsection{Models}
\subsubsection{Network Model} 
In this paper, we consider a wireless network, which can be either single-hop or multi-hop. In a single-hop network, there are a total of $N$ nodes, and any two nodes can communicate directly. In a multi-hop network, the network is divided into $M$ clusters, with each cluster consisting of a single-hop network. The $i^{th}$ cluster contains $N_i$ nodes. Nodes between clusters communicate through a routing protocol. Existing Byzantine fault-tolerant routing protocols \cite{curtmola2008bsmr, awerbuch2002demand} can be employed in our paper. Nodes are battery-powered and equipped with half-duplex communication devices. Each node is assigned a unique ID. At the beginning of the protocol, nodes are aware of the total number of nodes in the network. We adopt an asynchronous model, where message delays between nodes are unbounded. However, we assume that messages between honest nodes will eventually be delivered, a necessary condition for guaranteeing protocol security.

\subsubsection{Adversary Model} 
The nodes are categorized as either honest or Byzantine. Honest nodes strictly follow the protocol, while Byzantine nodes engage in arbitrary behaviors. For example, Byzantine nodes can arbitrarily prolong the delay between messages of two nodes and alter the order of message delivery. 
In a single-hop network, the maximum number of Byzantine nodes, $f$, is determined by the total number of nodes, following the relation $N = 3f + 1$. In a multi-hop network, for the $i^{th}$ cluster, the maximum number of Byzantine nodes, $f_i$, is determined by the total number of nodes in that cluster, following the relation $N_i = 3f_i + 1$.
Our work mainly focuses on designing the packet structure of asynchronous BFT consensus to better adapt to wireless networks, thereby enhancing efficiency without modifying the protocol's process and consequently not compromising safety or liveness. Additionally, we rely on the security of cryptographic primitives used in this paper.

\subsection{Preliminaries}
\subsubsection{Broadcast Protocols}
Asynchronous BFT consensus utilizes three broadcast protocols: reliable broadcast protocol (RBC)~\cite{cachin2005asynchronous, bracha1987asynchronous}, provable reliable broadcast protocol (PRBC)~\cite{guo2020dumbo}, or consistent broadcast protocol (CBC)~\cite{cachin2001secure}. We focus on their communication patterns to adapt these protocols from wired to wireless networks. We use ``x-to-y" to represent communication, where ``x" is the number of senders and ``y" is the number of receivers. Each protocol has several phases: RBC consists of $\mathsf{INITIAL}$, $\mathsf{ECHO}$, and $\mathsf{READY}$ (Both Bracha’s RBC~\cite{bracha1987asynchronous} and Cachin’s RBC~\cite{cachin2005asynchronous} can be illustrated by Fig.~\ref{fig:RBC}.); PRBC extends RBC by adding $\mathsf{DONE}$ phase (Fig.~\ref{fig:RBC}); CBC has $\mathsf{INITIAL}$, $\mathsf{ECHO}$, and $\mathsf{FINISH}$ (Fig.~\ref{fig:CBC}). In the $\mathsf{INITIAL}$ phase, a single node broadcasts its proposal using 1-to-N communication. In RBC and PRBC, the following two phases use N-to-N communication to broadcast votes.
In the $\mathsf{DONE}$ phase of PRBC, nodes collectively gather and combine threshold signature shares to form a complete signature. In contrast, CBC employs a two-phase approach: an N-to-1 communication round collects threshold signature shares, followed by a 1-to-N broadcast to distribute the final combined signature.

\subsubsection{ABA}
Asynchronous Byzantine Agreement (ABA) protocols rely on one or more broadcast mechanisms to achieve consensus on a proposed value. These protocols typically require a source of randomness, which can be implemented using either a local coin or a shared coin. Two prominent ABA protocols are Bracha’s ABA~\cite{bracha1984asynchronous} and Cachin’s ABA~\cite{cachin2000random}, both of which operate in a round-based manner. Figure~\ref{fig:Brachas-ABA} illustrates round $r$ of Bracha’s ABA, where each round comprises three phases. Each phase involves $N$ parallel Reliable Broadcast (RBC) instances for broadcasting votes, resulting in a message complexity of $\mathcal{O}(N^3)$ ($N$ RBC instances, each with complexity $\mathcal{O}(N^2)$) in wired networks. Figure~\ref{fig:Cachin'sABA} depicts round $r$ of Cachin’s ABA, which also has three phases but employs N-to-N communication and incorporates a common coin mechanism, generated through either threshold signatures or coin flipping. Nodes use this mechanism to decide whether to proceed to the next round or finalize a decision. This design achieves a reduced message complexity of $\mathcal{O}(N^2)$.

\subsubsection{Asynchronous BFT Consensus}

\begin{figure}[h] 
\centering
  \subfloat[HoneyBadgerBFT (Parallel RBCs and ABAs)]{  \includegraphics[width=0.47\linewidth]{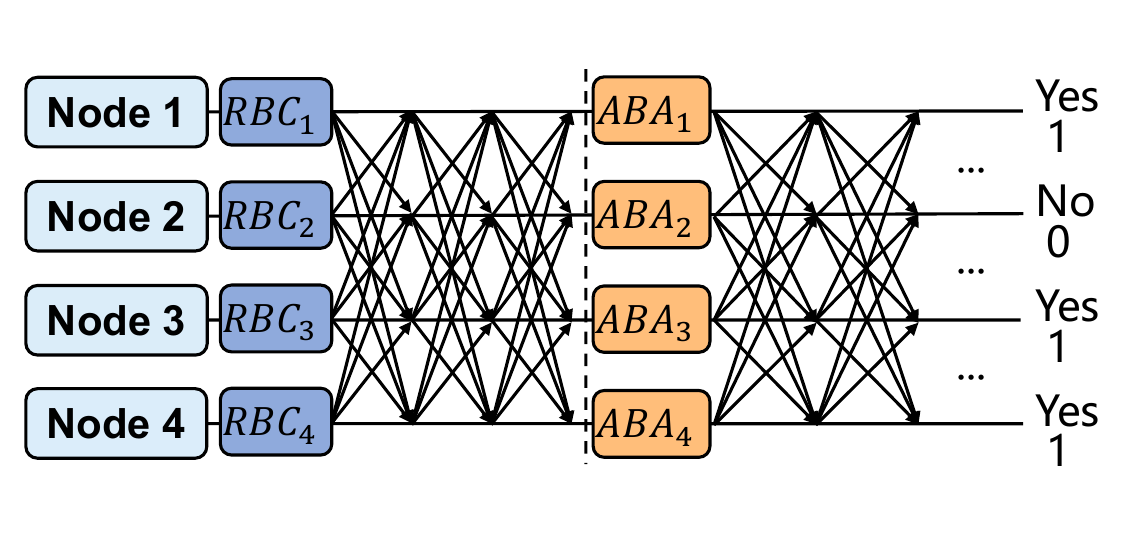}
\label{fig:HBBFT}}\hfill
  \subfloat[Dumbo (Parallel PRBCs and CBCs; Serial ABAs)]{  \includegraphics[width=0.47\linewidth]{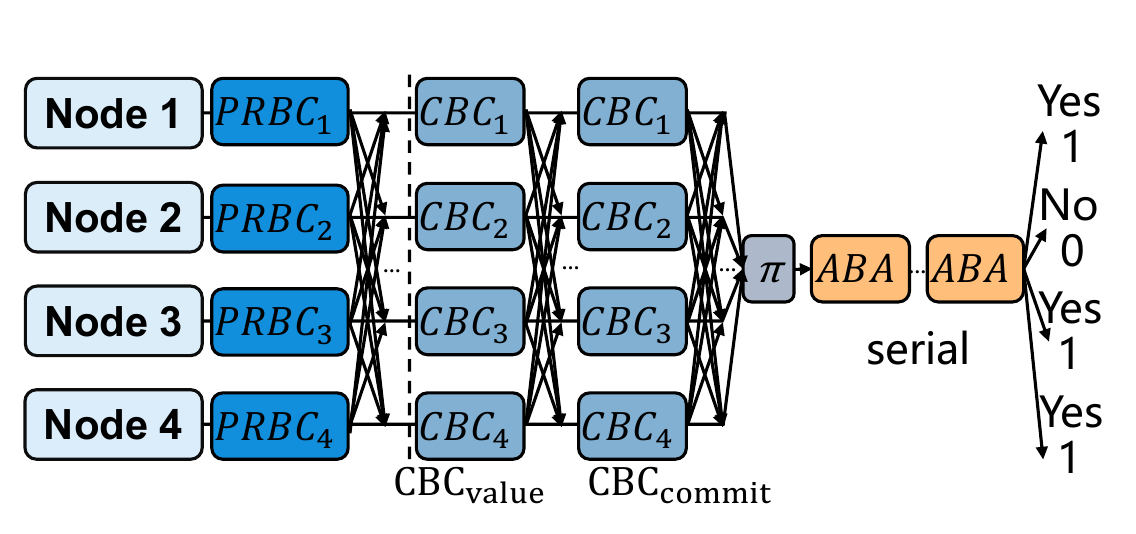}
\label{fig:Dumbo}}\hfill
  \caption{Diagrams of asynchronous BFT consensus}
  \label{fig:Flow_Consensus} 
\end{figure}

As shown in Fig.~\ref{fig:HBBFT}, HoneyBadgerBFT~\cite{miller2016honey} pioneers the practical implementation of asynchronous BFT consensus. It operates by concurrently executing N parallel instances of RBC and ABA. Each ABA instance is initiated by a single node but involves the participation of all nodes. The outcome of an ABA instance determines whether the proposal associated with the corresponding RBC instance is accepted.
BEAT~\cite{duan2018beat} builds upon HoneyBadgerBFT by integrating more efficient components. In this paper, we focus on BEAT's replacement of threshold signatures with threshold coin flipping. This modification does not alter the fundamental structure of HoneyBadgerBFT.
As shown in Fig.~\ref{fig:Dumbo}, to reduce the number of parallel ABA instances, Dumbo~\cite{guo2020dumbo} introduces a new architecture comprising $N$ parallel PRBC instances, two sets of $N$ parallel CBC instances (named $\mathsf{CBC_{value}}$ and $\mathsf{CBC_{commit}}$), and serial ABA instances. Compared to RBC, PRBC adds verification information to ensure that the final accepted proposal has been received by at least one honest node. The two sets of $N$ parallel CBC instances are used to unify the list of completed PRBC instances across all nodes. A global string $\pi$ is used to determine the execution order of subsequent ABAs. Each ABA corresponds to a set of proposals from the PRBC instances. The serial ABA instances execute until one ABA outputs 1, indicating acceptance of the corresponding proposal list. 

\section{Design of ConsensusBatcher}



\subsection{Network Congestion Caused by N Parallel Consensus Components}

The primary cause of network congestion in wireless networks is the increased channel contention caused by running $N$ parallel consensus components. Compared to a single consensus component, running $N$ parallel components amplifies message complexity by a factor of $N$. In wireless networks, nodes share a common channel. Time division multiple access (TDMA) and carrier sense multiple access (CSMA) are the two basic medium access control protocols. While TDMA can address channel contention in synchronous and partially synchronous networks, it is not applicable in asynchronous networks where message delays are unbounded. Therefore, CSMA becomes the only option for managing channel access. As a result, the amplified message complexity directly leads to increased channel contention by a factor of $N$, making limited communication resources the primary bottleneck for consensus. Thus, reducing channel contention is the most straightforward optimization strategy.

DMA (Direct Memory Access) improves data transfer efficiency by offloading data transmission tasks from the CPU, thereby freeing up computational resources. However, when running multiple parallel consensus components, the frequent transmission and reception of packets leads to DMA resource contention. This problem is particularly pronounced on single-core processor development boards, where such contention exacerbates transmission delays. On the other hand, cryptographic tools with high computational complexity—such as threshold encryption, threshold signatures, and threshold coin flipping—are critical to consensus protocols but involve time-consuming operations. This results in prolonged processing time for a single data packet. If multiple packets accumulate in the DMA buffer, it further extends processing time, further increasing consensus delays. Asynchronous BFT consensus protocols rely on timers to progress. Delays in message transmission and processing can cause timer timeouts, triggering repeated message transmission events and leading to network congestion. The use of DMA and cryptographic tools can indirectly contribute to network congestion.


\subsection{ConsensusBatcher}
To address the network congestion above, the core idea of ConsensusBatcher is to batch $N$ parallel consensus components by merging identical packet information across the components, thereby reducing the previously required $N$ channel access competitions to just one. To achieve this, ConsensusBatcher consists of a packet module, DMA module, and cryptographic module.

\subsubsection{Packet module}
The reliability mechanism ensures that messages are eventually delivered, which is critical for driving asynchronous BFT consensus. The two mainstream reliability mechanisms, Acknowledgment (ACK) and negative acknowledgment (NACK), are suited to different application scenarios, and selecting the appropriate one is essential for reducing network congestion. This choice is a key consideration for the packet module. For asynchronous wireless BFT consensus protocols, we opt for the NACK due to the following reasons. (1) In asynchronous BFT consensus, nodes advance to the next phase upon receiving a sufficient number of votes, regardless of sender acknowledgment. This decentralized decision-making mechanism aligns well with the NACK-based reliability mechanism. For example, in Bracha's RBC protocol, a node accepts a proposal after collecting $2f+1$ ready votes for that proposal. (2) Asynchronous BFT consensus primarily relies on a one-to-many communication paradigm, requiring a single message to be delivered to multiple nodes. In wireless networks with shared channels, ACK needs at least $N+1$ messages, while NACK requires only the initial message.

Batching $N$ consensus components requires careful design of the packet structure, ensuring that the partitioning adapts to upper-layer components and consensus protocols. Therefore, the packet module divides the packet payload into four parts: header, NACK, value, and signature.

\setlength{\fboxsep}{2pt}

\begin{itemize}
    \item \colorbox{color1}{Header}: The header contains fundamental packet metadata, including node identity, packet type, and multi-hop routing information.
    \item \colorbox{color2}{NACK}: The NACK is used for ensuring message delivery.
    \item \colorbox{color3}{Value}: The value is used to store proposals or the hash of a proposal.
    \item \colorbox{color4}{Signature}: The signature incorporates both public-key digital and threshold signatures.
\end{itemize}

\begin{figure}[!htb]
    \centering
    \includegraphics[width=0.8\linewidth]{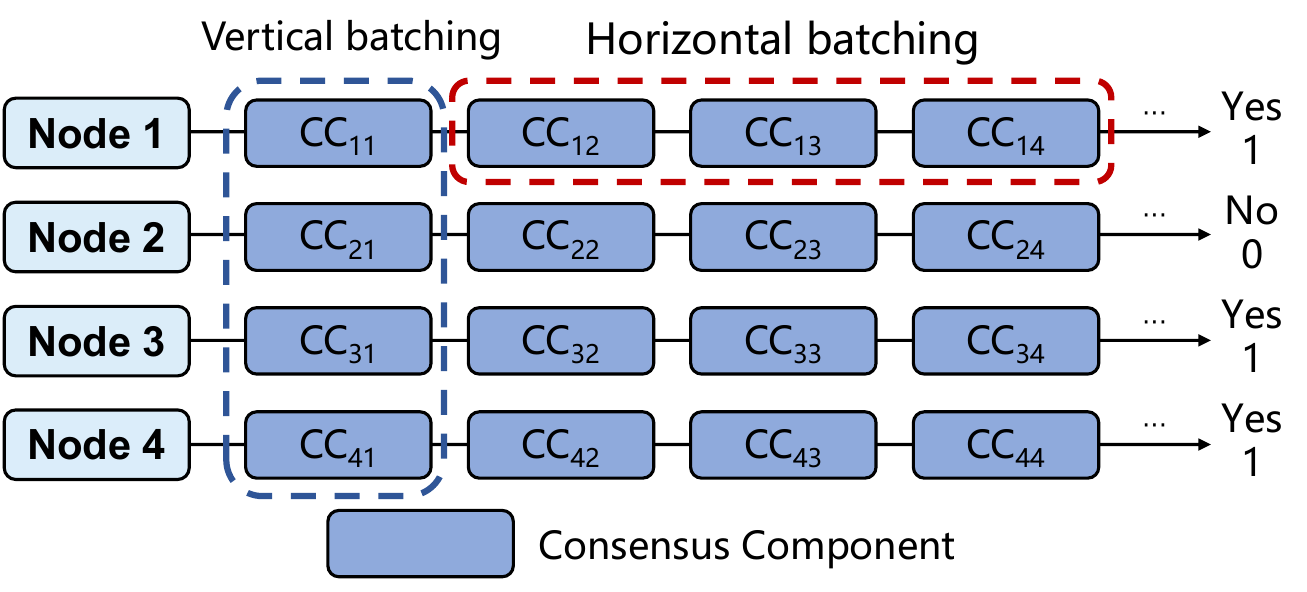}
    \caption{Schematic diagram of vertical and horizontal batching in ConsensusBatcher}
    \label{fig:ConsensusBatcher}
\end{figure}

Fig.~\ref{fig:ConsensusBatcher} provides a schematic representation of ConsensusBatcher, emphasizing its ability to batch $N$ consensus components both vertically and horizontally. Vertical batching aggregates messages from $N$ parallel consensus components within the same phase, allowing them to be processed collectively as a single communication unit. This consolidation effectively reduces the frequency of channel access contention, making vertical batching the principal optimization strategy in ConsensusBatcher. 
In contrast, horizontal batching combines messages from different phases of the same consensus component to harmonize execution speeds across its phases. For instance, in $N$ instances of RBC, some may be in the $\mathsf{ECHO}$ phase while others are in the $\mathsf{READY}$ phase. By applying horizontal batching, two channel access events are condensed into one, further streamlining communication. 
Additionally, the packet module incorporates optimization techniques tailored for integration with higher-layer protocols. The design and interaction of this module with consensus components are analyzed in Section~\ref{sec:CB:cc}, while Section~\ref{sec:consensus} explores its role within consensus protocols.

\subsubsection{DMA Module} 
%
Building upon the DMA half-interrupt mechanism, we introduce a packet alignment strategy integrated into the DMA module to further improve efficiency. This mechanism establishes a relationship between the DMA buffer size and the protocol packet length. Specifically, the buffer size is configured to align with the maximum packet length defined by the consensus protocol, denoted as $2D$. The alignment mechanism enforces that all packet lengths conform to either the full buffer size $2D$ or fall within the range $[D, 2D)$. Packets shorter than $D$ are padded to reach this threshold. 
Packets of length $2D$ prompt a full interrupt, bypassing the half-interrupt process and triggering an immediate reset of the DMA buffer. For packets within the range $[D, 2D)$, a half interrupt is generated, followed by a buffer reset. Consequently, packets do not accumulate in the DMA buffer, and a CPU interrupt is invoked immediately for processing. This approach minimizes processing latency and indirectly alleviates network congestion.

\subsubsection{Cryptographic Module}
The cryptographic module primarily provides lightweight implementations of public-key digital signatures and threshold cryptography. Using the lightweight MIRACL library\footnote{https://github.com/miracl}, we implement threshold encryption, threshold signatures, and threshold coin flipping. For public-key digital signatures, we utilize the existing micro-ecc cryptographic library\footnote{https://github.com/kmackay/micro-ecc}. Public-key digital signatures and threshold cryptography exhibit variations in signature size and computation time depending on the elliptic curve used, with these differences being particularly significant in resource-constrained environments. Minimizing the size of public-key digital and threshold signatures allows for more space for other data, enabling more consensus component instances to be batched and improving overall consensus efficiency. In this module, we delve into the performance characteristics of public-key digital signatures and threshold cryptography schemes. We specifically examine signature size and computation time metrics across a range of elliptic curves. The results of our experiments and a detailed discussion are provided in Section~\ref{sec:evaluation}.

\subsection{Consensus Components}
\label{sec:CB:cc}
Next, we introduce how ConsensusBatcher is used for batching $N$ parallel consensus components, including three broadcast protocols and two ABA protocols. We also discuss the technical challenges encountered during both vertical and horizontal batching.

\subsubsection{Broadcast Protocol}

The asynchronous BFT consensus uses two types of RBC: Cachin's RBC \cite{cachin2005asynchronous} and Bracha's RBC \cite{bracha1987asynchronous}. Cachin's RBC divides the proposal into N blocks, sending different blocks to different nodes, which requires $N-1$ broadcasts and underutilizes the broadcast capabilities of wireless channels. In contrast, Bracha's RBC broadcasts the entire proposal in a single operation, making it more efficient for wireless environments. Therefore, in this paper, the term RBC specifically refers to Bracha's RBC.

We first apply ConsensusBatcher to perform vertical batching for the three broadcast protocols. Vertical batching merges the same phase across $N$ broadcast protocol instances. There are three key aspects to consider during vertical batching. 

\begin{figure}[!h] 
\centering
  \subfloat[Packet structure of $N$ parallel RBC instances]{
   \includegraphics[width=0.95\linewidth]{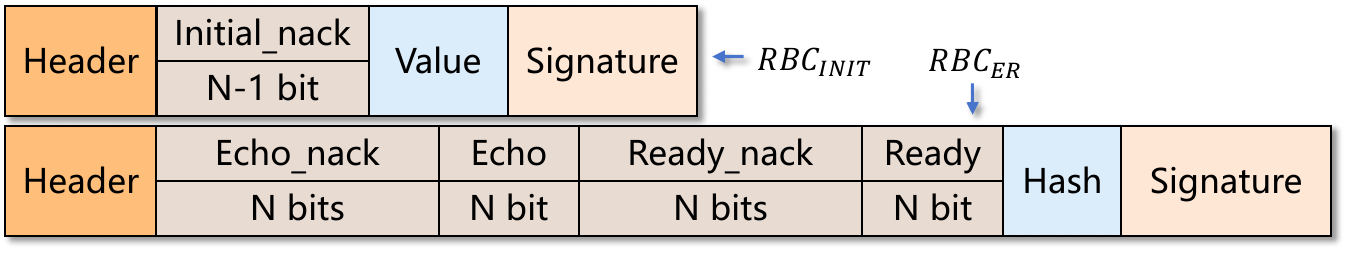}
\label{fig:packet_RBC}}\\
  \subfloat[Packet structure of $N$ parallel CBC instances]{
    \includegraphics[width=0.95\linewidth]{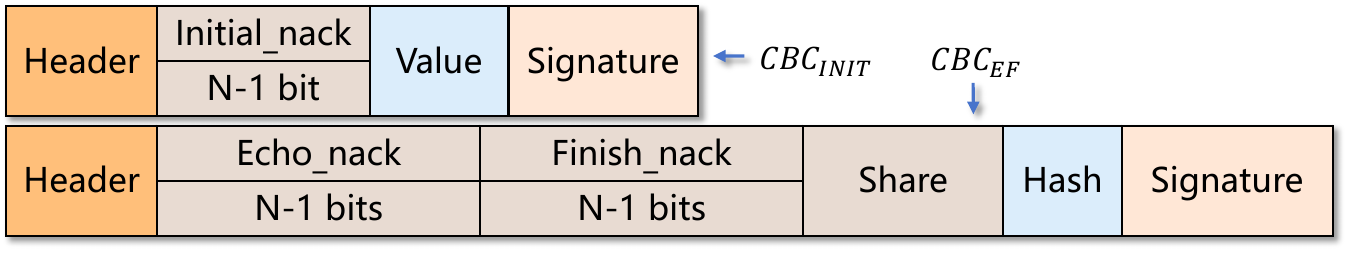}
\label{fig:packet_CBC}}\\
  \subfloat[Packet structure of $N$ parallel PRBC instances]{
    \includegraphics[width=0.95\linewidth]{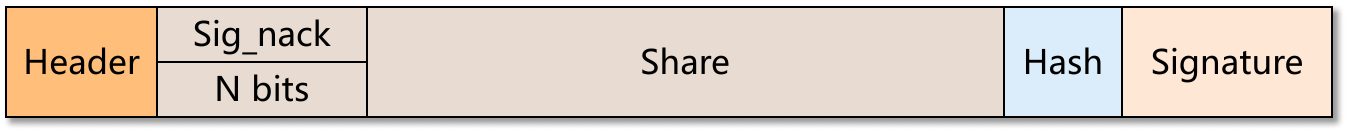}
\label{fig:packet_PRBC}}
  \caption{Packet structure of broadcast protocols}
  \label{fig:packet_BC} 
\end{figure}

\begin{itemize}
    \item We observe that the $\mathsf{INITIAL}$ phase in all three broadcast protocols involves large proposals that span multiple packets. Therefore, for this phase, vertical batching applies only to its NACK, i.e. $\mathsf{Initial\_nack}$ in $\mathsf{RBC_{INIT}}$ (Fig.~\ref{fig:packet_RBC}) and $\mathsf{CBC_{INIT}}$ (Fig.~\ref{fig:packet_CBC}). 
    
    \item For phases other than the $\mathsf{INITIAL}$ phase, batching requires identifying each proposal using its ID. As shown in $\mathsf{RBC_{ER}}$ (Fig.~\ref{fig:packet_RBC}) and $\mathsf{CBC_{EF}}$ (Fig.~\ref{fig:packet_CBC}), we use a hash part to store the hash of each of the $N$ proposals, enabling their identification. 

    \item The NACK in vertical batching can be further compressed, reducing space complexity from $\mathcal{O}(N^2)$ to $\mathcal{O}(N)$. For example, in the $\mathsf{ECHO}$ phase of RBC, the NACK for a single RBC instance consists of $N-1$ bits, where each bit corresponds to a node and indicates the reception of the echo from the corresponding node. However, when vertical batching $N$ RBC instances, as shown in $\mathsf{RBC_{ER}}$ (Fig.~\ref{fig:packet_RBC}), the NACK requires $N(N-1)$ bits, occupying significant packet space. To address this, ConsensusBatcher uses $N$ bits, where each bit represents one RBC instance and indicates whether the RBC instance has received $2f+1$ echo messages. This optimization reduces the space complexity of the NACK from $\mathcal{O}(N^2)$ to $\mathcal{O}(N)$.
\end{itemize}

Next, we further apply ConsensusBatcher for horizontal batching of the three broadcast protocols. Due to the large size of proposals in the $\mathsf{INITIAL}$ phase, horizontal batching focuses on merging phases other than $\mathsf{INITIAL}$. As shown in Fig.~\ref{fig:packet_RBC} and Fig.~\ref{fig:packet_CBC}, the packet structure for $N$ parallel RBC instances and $N$ parallel CBC instances is divided into two types. Since the $\mathsf{DONE}$ phase in PRBC requires threshold signatures, which occupy significant packet space, we batch it separately, as shown in Fig.~\ref{fig:packet_PRBC}.

\begin{mybox}[boxsep=0pt,
 boxrule=1pt,
 left=4pt,
 right=4pt,
  top=4pt,
  bottom=4pt,
 ]
 \textbf{Technical Challenge~I:}  
In the scenarios of RBC in Bracha's ABA and $\mathsf{CBC_{commit}}$ in Dumbo, the proposals in these broadcast protocols are small, directly using the normal packet structures waste bandwidth.
\end{mybox}

\begin{figure}[h] 
\centering
  \subfloat[Packet structure of $N$ parallel RBC instances with small proposals]{
    \includegraphics[width=0.95\linewidth]{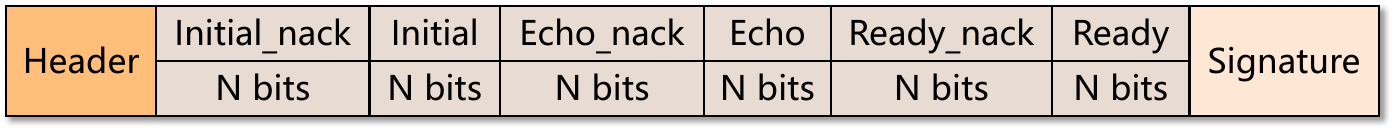}
\label{fig:RBC-small}}\\
  \subfloat[Packet structure of $N$ parallel CBC instances with small proposals]{
    \includegraphics[width=0.95\linewidth]{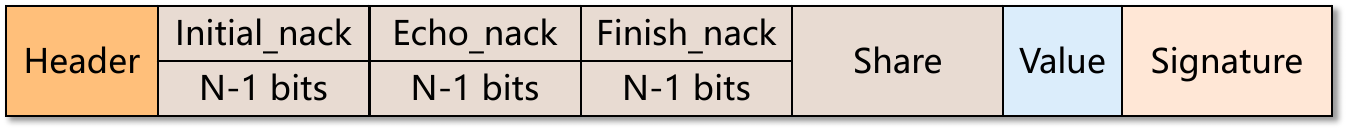}
\label{fig:CBC-small}}
  \caption{Packet structure of $N$ parallel broadcast protocols with small proposal size}
  \label{fig:3PCBC} 
\end{figure}

When the size of proposals in RBC or CBC is small, ConsensusBatcher can perform horizontal batching for $N$ parallel instances further. The packet structures shown in Fig.\ref{fig:packet_RBC} and Fig.\ref{fig:packet_CBC} have already undergone vertical batching by ConsensusBatcher, so horizontal batching is performed on these packet structures. This involves integrating the $\mathsf{INITIAL}$ phase with the other two phases. In Bracha's ABA, $N$ parallel RBC instances are executed, where the proposal broadcast by RBC has only three possible values: 1, 0, and $\perp$. Thus, only two bits are needed to represent all possible proposals. As a result, $N$ parallel RBC instances require $2N$ bits, corresponding to $\mathsf{Initial\_nack}$ and $\mathsf{Initial}$, as shown in Fig.~\ref{fig:RBC-small} referred to as RBC-small. In Dumbo, the proposals in $N$ parallel $\mathsf{CBC_{commit}}$ instances are node ID lists of length $2f+1$. Here, we can use $N$ bits to denote one proposal, as illustrated in Fig.~\ref{fig:CBC-small}, referred to as CBC-small. Additionally, if the length of $N$ bits exceeds the length of a hash, using a hash is more space-efficient for the packet.

\subsubsection{ABA}
In asynchronous BFT consensus, ABA are implemented in two ways: parallel (e.g., HoneyBadgerBFT) and serial (e.g., Dumbo).

\begin{mybox}[boxsep=0pt,
 boxrule=1pt,
 left=4pt,
 right=4pt,
  top=4pt,
  bottom=4pt,
 ]
 \textbf{Technical Challenge~II:} 
Using standard packets or packets optimized solely for RBC still incurs message overhead when implementing Bracha's ABA, whether in parallel or serial configurations. Furthermore, designing a single packet format that optimally supports both parallel and serial ABA implementations is challenging.
\end{mybox}

\begin{figure}[h] 
\centering
  \subfloat[Packet structure of $k$ parallel Bracha's ABA instances]{
   \includegraphics[width=0.95\linewidth]{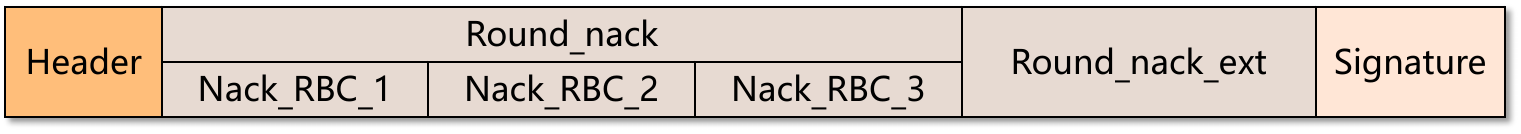}
\label{fig:Bracha's ABA packet}}\\
  \subfloat[Packet structure of $k$ parallel Cachin's ABA instances]{
    \includegraphics[width=0.95\linewidth]{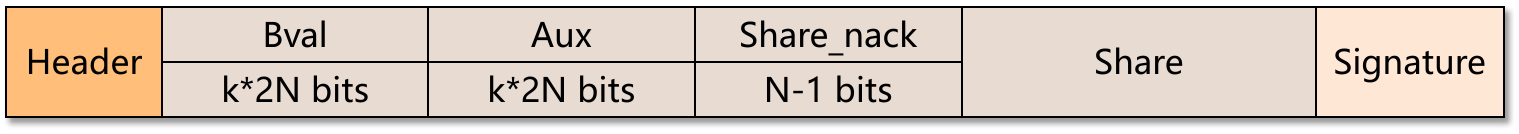}
\label{fig:Cachin's ABA packet}}
  \caption{Packet structure of $k$ parallel ABA instances}
  \label{fig:ABA packet structure} 
\end{figure}

For parallel Bracha’s ABA instances, since each ABA contains three-phase RBCs, we can first perform horizontal batching by merging the three phases. As shown in Fig.~\ref{fig:Bracha's ABA packet}, $\mathsf{Nack\_RBC\_1}$, $\mathsf{Nack\_RBC\_2}$, and $\mathsf{Nack\_RBC\_3}$ each consist of the $\mathsf{NACK}$ defined in Fig.~\ref{fig:RBC-small}, representing the three phases in Bracha’s ABA, respectively. Next, we perform vertical batching by merging multiple Bracha’s ABA instances. To achieve this, we assign $\mathsf{Round\_nack\_ext}$ to include multiple Bracha’s ABA instances, with a variable length depending on the number of ABA instances processed in parallel. 
For serial Bracha’s ABA instances, we only perform horizontal batching on both three-phase RBCs. The packet is shown in Fig.~\ref{fig:Bracha's ABA packet} without $\mathsf{Round\_nack\_ext}$.


\begin{mybox}[boxsep=0pt,
 boxrule=1pt,
 left=4pt,
 right=4pt,
  top=4pt,
  bottom=4pt,
 ]
 \textbf{Technical Challenge~III:}  
Parallel instances of Cachin's ABA use different common coins for the same round, which fails to leverage the broadcast advantage of wireless networks.
\end{mybox}

In wired networks, if different ABA instances use the same common coin in the same round, it leads to security issues. Byzantine nodes can exploit timing differences in ABA execution to obtain the common coin from faster-running ABA instances. By manipulating the order of votes from $\mathsf{BVAL}$ and $\mathsf{AUX}$, phases in Cachin’s ABA, arriving at honest nodes, Byzantine nodes can prevent different ABAs from completing in the same round, thereby reducing overall consensus efficiency. However, in wireless networks, nodes share the channel. Even if Byzantine nodes gain access to the common coin in advance, the broadcasting nature of wireless networks, combined with the fact that votes within the ConsensusBatcher are bound together, prevents them from manipulating the order of votes. 


For parallel Cachin’s ABA instances, to merge the votes of the $\mathsf{BVAL}$ and $\mathsf{AUX}$ phases, we can first perform horizontal batching by merging the three phases in Cachin’s ABA, corresponding to $\mathsf{Bval}$, $\mathsf{Aux}$, and $\mathsf{Share\_nack}$ in Fig.~\ref{fig:Cachin's ABA packet}. Next, we perform vertical batching by merging multiple Cachin’s ABA instances. To achieve this, we expand $\mathsf{Bval}$ and $\mathsf{Aux}$ $k$ times, merging $k$ Cachin’s ABA instances. Since the same round of $k$ ABA instances can share a common coin, $\mathsf{Share\_nack}$ does not need to be expanded $k$ times.
For serial Cachin’s ABA instances, we only perform horizontal batching. This corresponds to Fig.~\ref{fig:Cachin's ABA packet}, where the expansion factor $k$ is removed from $\mathsf{Bval}$ and $\mathsf{Aux}$.


\subsubsection{Analysis}


To accurately represent the number of messages in the protocol, we use ``message overhead" instead of ``message complexity," as message overhead can precisely reflect the exact number of messages. As shown in Table~\ref{tab:mc}, we analyze the message overhead per node in various $N$-component parallel protocols. For wired networks, broadcasting a message to $N$ nodes necessitates $N$ transmissions. In wireless networks, leveraging the shared channel, this overhead is reduced to a single transmission, setting the baseline. 
Consequently, for consensus components with N-to-N communication, the number of channel contention instances in wired networks can be reduced by a factor of $\mathcal{O}(N)$ in the wireless baseline. For CBC protocols, which inherently employ 1-to-N and N-to-1 communication, the wireless baseline reduces the total message overhead from $3(N-1)$ to $N+1$.




\begin{table}[h]
\centering
\caption{Comparison of message overhead per node in an $N$-component parallel protocol}
\label{tab:mc}
\begin{threeparttable}
\begin{adjustbox}{width=0.95\linewidth} 
\tabcolsep=0.2cm
    \setlength{\tabcolsep}{3pt}{
    \begin{tabular}{l c c c}
        \toprule[1pt]
                & Wired Network & \makecell[c]{Baseline \\ Wireless Network} & \makecell[c]{\textbf{ConsensusBatcher} \\ \textbf{Wireless Network}}  \\ 
        \midrule[0.5pt]
        RBC & $(N-1)(1+2N)$ & $1+2N$ & $1+2$   \\ 
        CBC & $3(N-1)$ & $1+(N-1)+1$ & $1+1+1$   \\
        PRBC & $(N-1)(1+3N)$ & $1+3N$ & $1+3$   \\
        Bracha's ABA & $3N(N-1)(1+2N)$ & $3N(1+2N)$ & $3(1+2)$   \\
        Cachin's ABA & $3N(N-1)$ & $3N$ & $3$   \\
        \bottomrule[1pt]
    \end{tabular}
    }
    \end{adjustbox}
\end{threeparttable}
\end{table}

The goal of ConsensusBatcher is to reduce the message overhead of $N$ parallel consensus components to that of a single consensus component. For RBC, PRBC, and Cachin’s ABA, the message overhead per node can be reduced from $\mathcal{O}(N^2)$ (in wired networks) to $\mathcal{O}(1)$ (in ConsensusBatcher wireless networks). For CBC, the message overhead per node can be reduced from $\mathcal{O}(N)$ (in wired networks) to $\mathcal{O}(1)$ (in ConsensusBatcher wireless networks). For Bracha’s ABA, the message overhead per node can be reduced from $\mathcal{O}(N^3)$ (in wired networks) to $\mathcal{O}(1)$ (in ConsensusBatcher wireless networks). In addition, ConsensusBatcher not only reduces message overhead but also lowers communication overhead in two ways. Firstly, each message requires a public-key digital signature, so reducing the message overhead also decreases the number of public-key digital signatures required. Secondly, NACK in ConsensusBatcher can convey more information, reducing their overhead from $\mathcal{O}(N^2)$ to $\mathcal{O}(N)$, as exemplified by $\mathsf{Echo\_nack}$ in Fig.~\ref{fig:packet_RBC}. Therefore, ConsensusBatcher effectively reduces the overhead of $N$ parallel consensus components in both message overhead and communication overhead.

\section{Asynchronous Wireless BFT  Consensus}
\label{sec:consensus}
In this section, we discuss how to utilize ConsensusBatcher-based consensus components to implement asynchronous BFT consensus\footnote{Both BEAT and Dumbo are families of consensus protocols. We select the most fundamental and critical consensus from each as an example, namely BEAT0 and Dumbo2.}.

\subsection{Single-Hop}
The consensus components can be further optimized before being applied to consensus protocols. First, the conditions for linking different consensus components vary across protocols, requiring adjustments to address these differences. Second, additional optimizations can be applied during the integration of consensus components to reduce network congestion. In detail, we refer to the ConsensusBatcher-optimized consensus components as ConsensusBatcher-X, such as ConsensusBatcher-RBC.

\begin{figure}[h] 
\centering
  \subfloat[Wireless HoneyBadgerBFT]{  \includegraphics[width=0.47\linewidth]{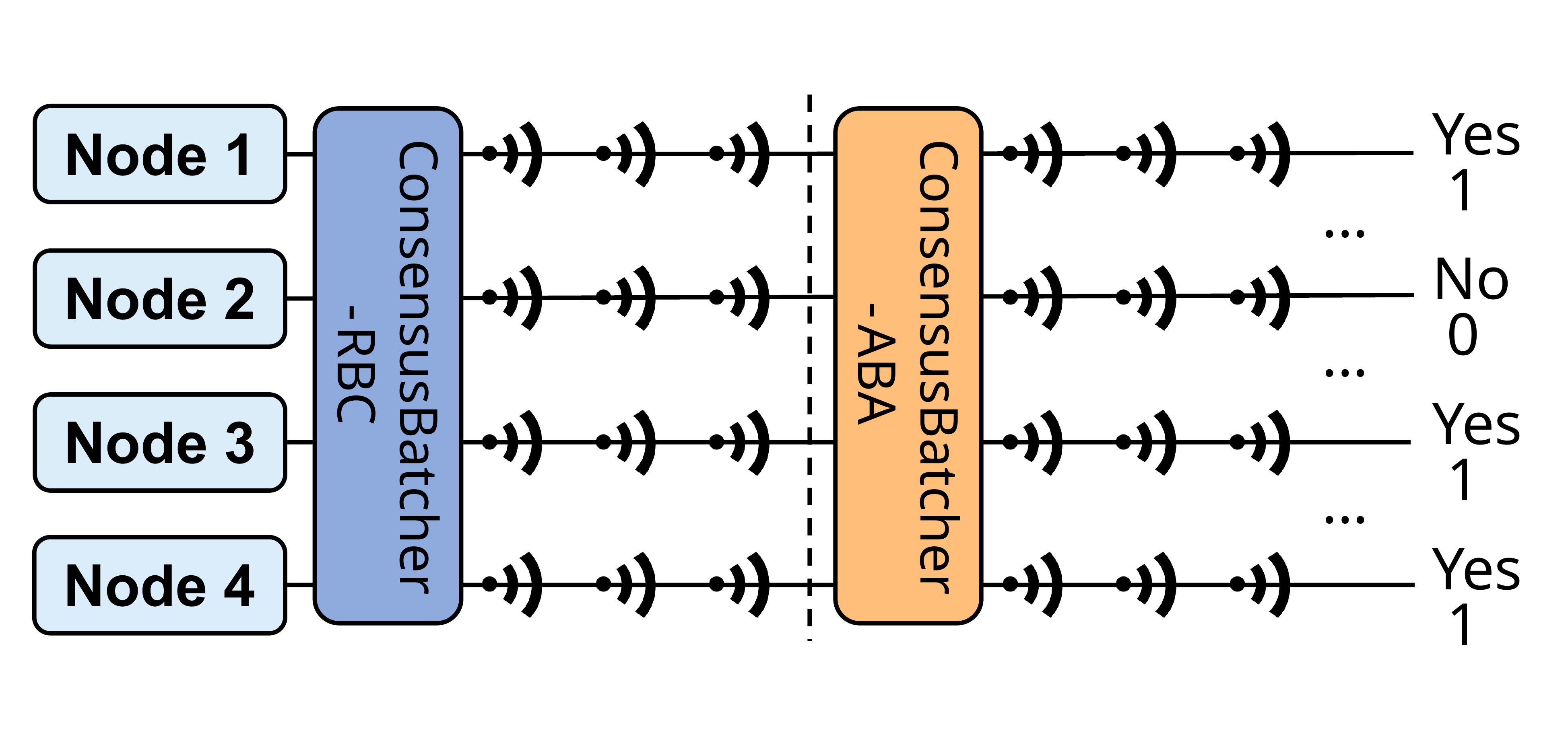}
\label{fig:wirelessHBBFT}}\hfill
  \subfloat[Wireless Dumbo]{  \includegraphics[width=0.47\linewidth]{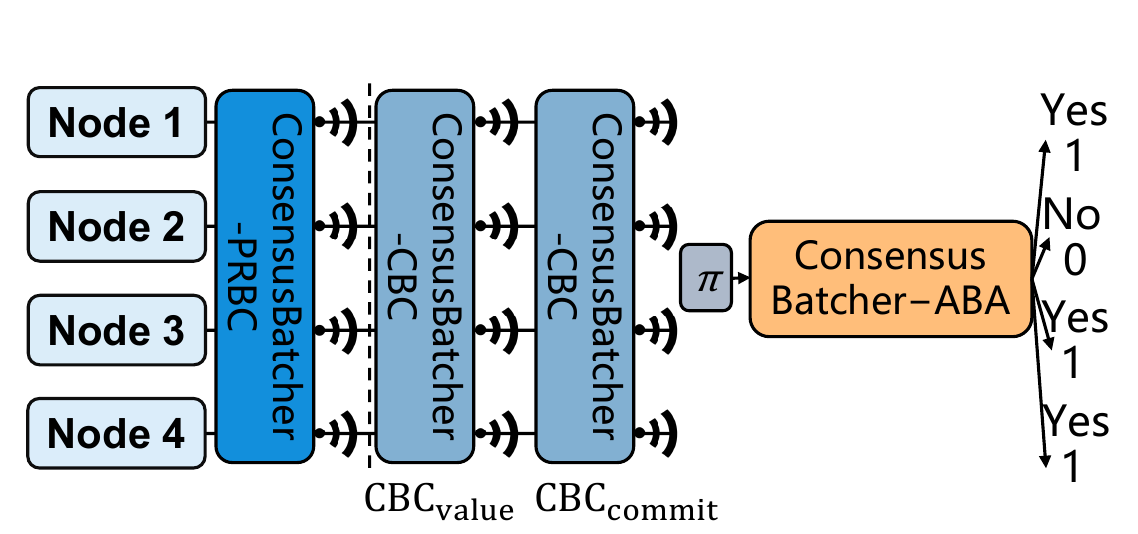}
\label{fig:wirelessDumbo}}\hfill
  \caption{Diagrams of asynchronous wireless BFT consensus}
  \label{fig:Async_Consensus} 
\end{figure}

In wireless HoneyBadgerBFT (as shown in Fig.~\ref{fig:wirelessHBBFT}), ABA instances in ConsensusBatcher-ABA must start simultaneously, unlike in wired networks where they can begin at different times. This synchronization prevents Byzantine nodes from gaining early access to the common coin, ensuring liveness. Nodes vote 1 for the ABA instances corresponding to the $2f+1$ fastest-completing RBC instances and 0 for the others, reducing the risk of Byzantine nodes delaying RBC completion. As a result, once $2f+1$ RBC instances in ConsensusBatcher-RBC are completed, all ABA instances in ConsensusBatcher-ABA are triggered concurrently.
Wireless BEAT retains the structure of wireless HoneyBadgerBFT but replaces threshold signatures with threshold coin flipping. This modification introduces additional verification data in the $\mathsf{SHARE}$ phase to validate share values.

In wireless Dumbo (as shown in Fig.~\ref{fig:wirelessDumbo}), after $2f+1$ PRBC instances in ConsensusBatcher-PRBC are completed, the protocol transitions to two sets of $N$ parallel CBC instances, i.e. ConsensusBatcher-CBC. These two sets called $\mathsf{CBC_{value}}$ and $\mathsf{CBC_{commit}}$, are used to synchronize the PRBC proposals. Once $2f+1$ parallel $\mathsf{CBC_{value}}$ instances are completed, the protocol moves to $N$ parallel $\mathsf{CBC_{commit}}$ instances. After $2f+1$ $\mathsf{CBC_{commit}}$ instances are finished, a global $\pi$ value is determined, which establishes the execution order for the subsequent serial ABA instances. We use ConsensusBatcher-ABA to replace serial ABA instances because ConsensusBatcher-ABA transforms multiple parallel ABA instances into multiple rounds of parallel ABA, which can be applied to serial ABA execution. During the serial ABA execution, it is essential to prevent premature calculation of share values for subsequent ABA instances. This ensures that Byzantine nodes cannot gain early access to the common coin, which could compromise liveness.


\subsection{Multi-Hop}
The architecture of multi-hop wireless networks greatly impacts the performance of asynchronous wireless BFT consensus. A naive approach involves all nodes participating in consensus, relying on routing protocols for communication. However, this leads to excessive routing information in packets, reducing space for proposals, and requires numerous parallel consensus component instances. With fixed packet parallelism $D$, increasing the network size $N$ raises message complexity and worsens congestion.
To address these issues, we divide the network into clusters, each functioning as a single-hop network. This clustering reduces routing overhead, improves packet efficiency, and ensures more efficient execution of the consensus protocol.

\begin{figure}[!htb]
    \centering
    \includegraphics[width=0.8\linewidth]{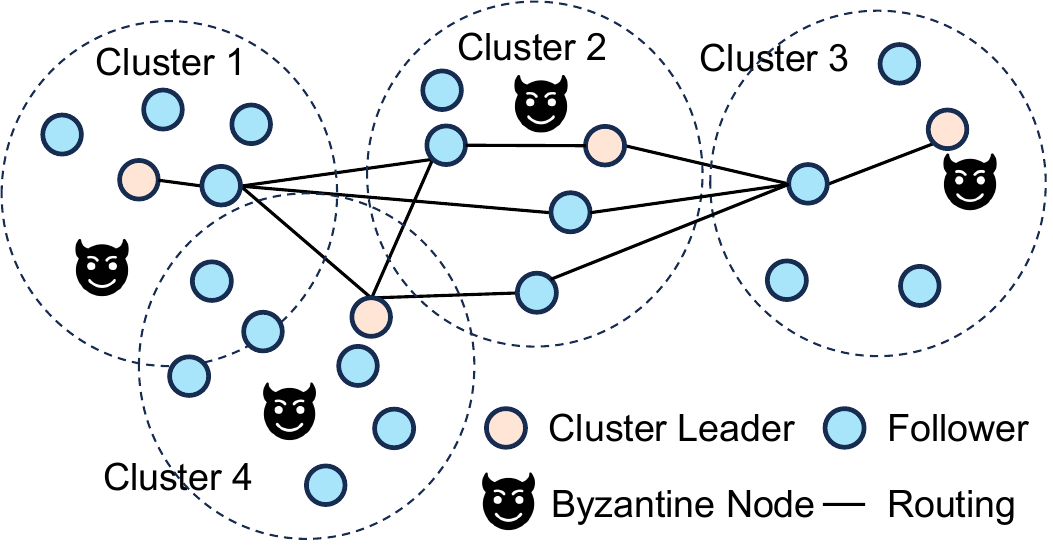}
    \caption{Multi-hop network topology}
    \label{fig:multi-hop network}
\end{figure}

Figure~\ref{fig:multi-hop network} illustrates a multi-hop network divided into four clusters, with each node belonging to one cluster. A two-phase consensus approach is used: local consensus within clusters and global consensus across clusters, akin to blockchain sharding~\cite{li2022jenga, hong2023prophet}. Local consensus is conducted in parallel within clusters without interference. Once completed, a changeable cluster leader from each cluster, carrying the cluster's proposals and signatures, is randomly selected and participates in the global consensus, which orders all clusters' proposals. Global consensus requires routing but involves far fewer nodes than routing across the entire network.
Local consensus ensures safety and liveness as long as Byzantine nodes remain under one-third of the total nodes. Global consensus may be controlled by Byzantine nodes. However, other nodes within the local consensus can detect erroneous proposals and change the wrong cluster leader. Ultimately, global consensus can ensure that fewer than one-third of the nodes are Byzantine, thus guaranteeing both safety and liveness.


\subsection{Asynchronous Wireless BFT  Consensus Testbed}
\label{sec:consensus:testbed}
\begin{figure}[h] 
\centering
  \subfloat[]{ \includegraphics[width=0.47\linewidth]{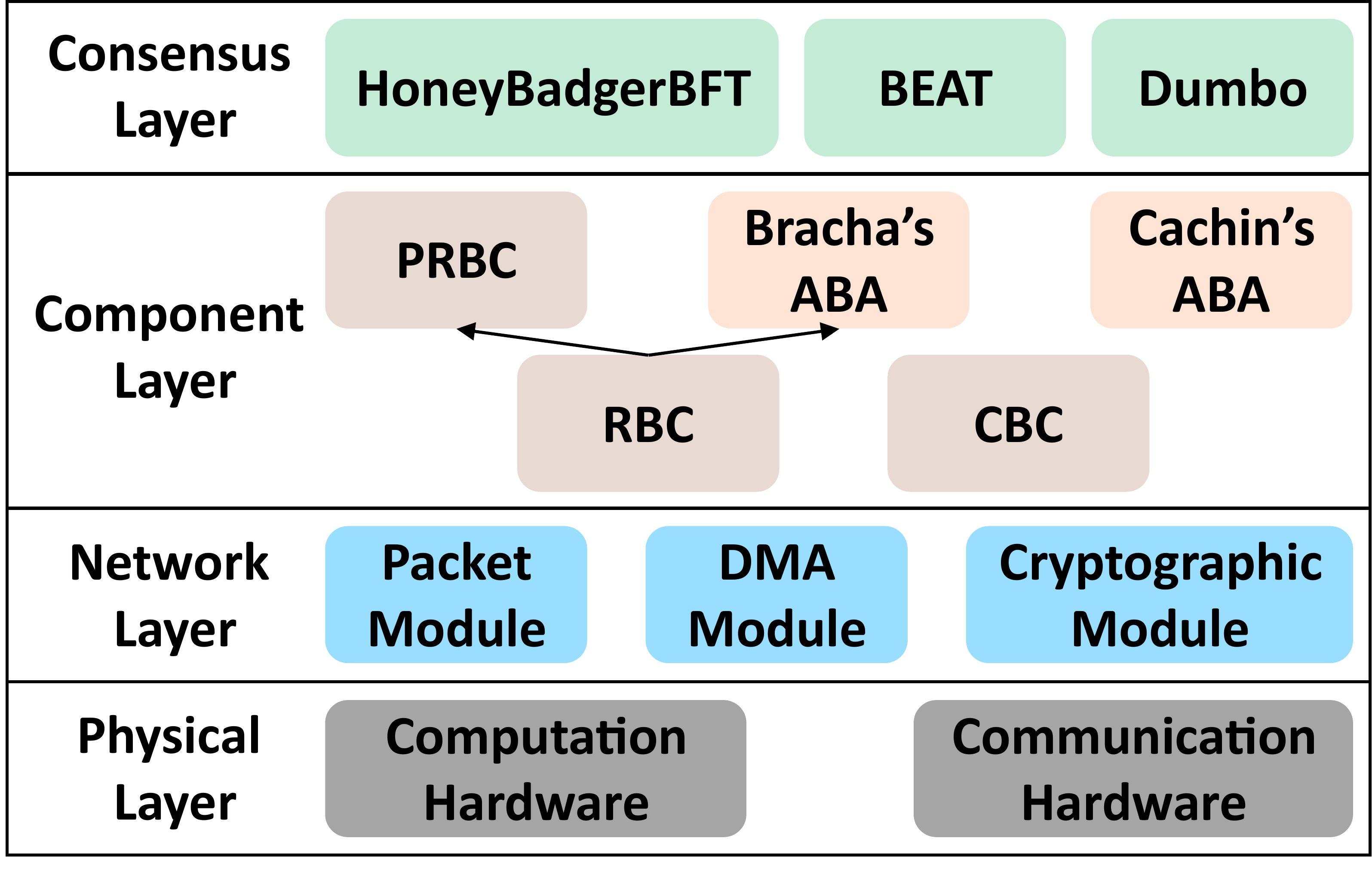}
\label{fig:architecture}}\hfill
  \subfloat[]{  \includegraphics[width=0.47\linewidth]{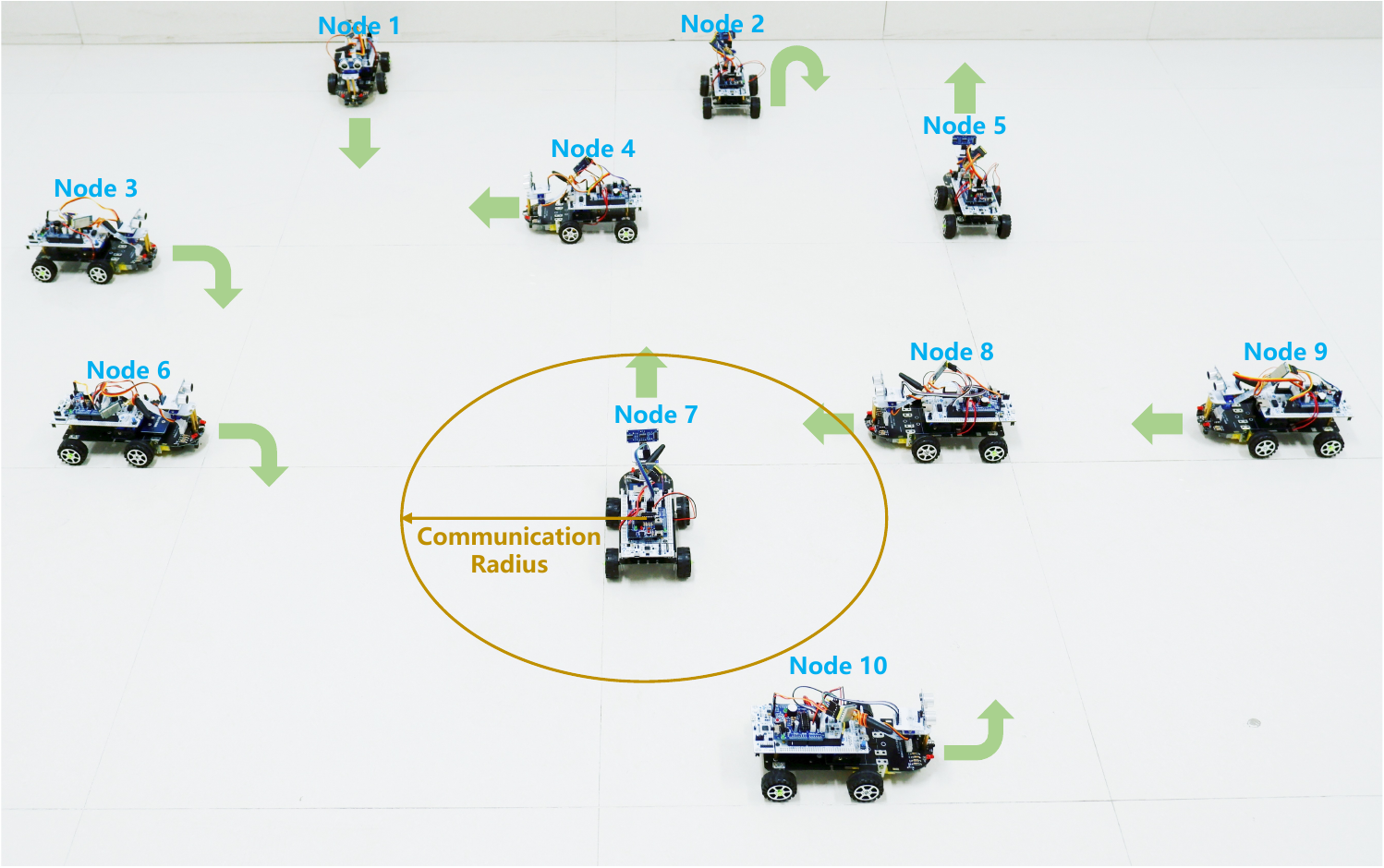}
\label{fig:carswarm}}\hfill
  \caption{Asynchronous wireless BFT consensus testbed. (a)Asynchronous wireless BFT consensus testbed architecture. The black arrows indicate the support relations. (b) Smart car swarm}
  \label{fig:testbed} 
\end{figure}

As shown in Fig.~\ref{fig:architecture}, our testbed adopts a four-layer structure: the physical layer, network layer, component layer, and consensus layer. This architecture features a modular design, where the four layers operate independently while maintaining interconnectivity. The code for the network layer and above has been open-sourced, enabling developers and researchers to perform further development based on it.

The physical layer provides the fundamental communication and computational capabilities. Our testbed does not impose specific constraints on the physical layer, as standard wireless network hardware, such as those used for Wi-Fi, Bluetooth, or LoRa, is sufficient to meet our requirements. While we employ LoRa and STM32F767 in our evaluation, other combinations, including Wi-Fi and Arduino or Bluetooth and Raspberry Pi, are also compatible with our testbed.
Fig.~\ref{fig:carswarm} depicts a test scenario involving ten intelligent vehicles moving randomly, with green arrows indicating their directions. Low-power antennas are used to limit the transmission range to 1 meter.

The component layer and the consensus layer provide the fundamental components of asynchronous BFT consensus and classic asynchronous BFT consensus protocols. This design facilitates secondary development, enabling developers to concentrate on upper-layer protocol design without being concerned with the implementation of lower-layer modules. The component layer provides two key types of components: broadcast protocols and ABA. For broadcast protocols, we offer implementations of RBC, PRBC, and CBC, as well as optimized versions for small broadcast data sizes, namely RBC-small and CBC-small. For ABA, we implement both Bracha’s ABA and Cachin’s ABA, supporting $N$ parallel and $N$ serial scenarios. Additionally, for parallel Cachin’s ABA, we provide an alternative implementation using threshold coin flipping instead of threshold signatures. The consensus layer builds on these consensus components to provide consensus algorithms. Based on two different ABA constructions, namely Bracha’s ABA and Cachin’s ABA, this layer implements two variations of HoneyBadgerBFT and two variations of Dumbo. Along with BEAT, this layer deploys a total of five distinct consensus protocols. Furthermore, for each of these five consensus protocols, we provide implementations for both single-hop and multi-hop network scenarios. 

\section{Evaluation}
\label{sec:evaluation}
This section evaluates the performance of asynchronous wireless BFT consensus using the testbed described in Section~\ref{sec:consensus:testbed}. We focus on three main aspects: cryptographic tools, consensus components, and consensus. Our test comprises 33,844 lines of C code. Each sample point is based on 30 measurements, and the mean value is calculated. Our evaluation addresses the following key questions:
\begin{itemize}
    \item What are the signature lengths and computation time of different implementations of cryptographic tools, and how do they impact consensus latency and throughput?
    \item What is the latency of consensus components, including broadcast protocols and ABA?
    \item What are the latency and throughput of the three asynchronous BFT consensus protocols in single-hop and multi-hop wireless networks? Compared to baseline consensus protocols, how much improvement in latency and throughput do consensus protocols based on ConsensusBatcher achieve?
\end{itemize}

\subsection{Cryptographic Tools}
\label{sec:evaluation:CT}

\begin{figure}[h] 
\centering
    \includegraphics[width=0.75\linewidth]{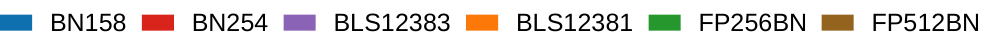}
    \label{fig:legend}
  \\
  \subfloat[The latency of basic operations in threshold signatures]{
    \includegraphics[width=0.47\linewidth]{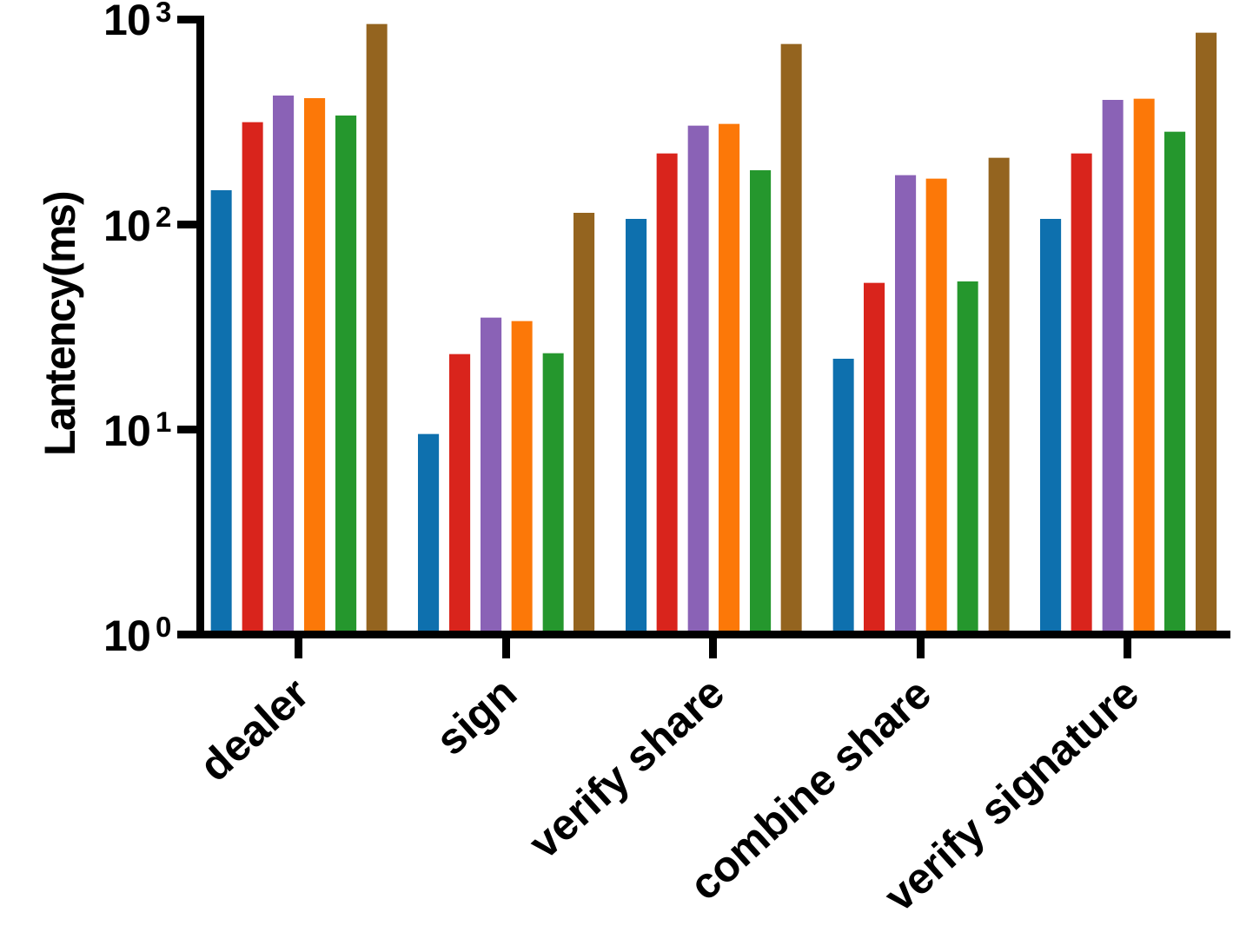}
    \label{fig:threshold_signature}
  }\hfill
  \subfloat[The latency of basic operations in threshold coin flipping]{
    \includegraphics[width=0.47\linewidth]{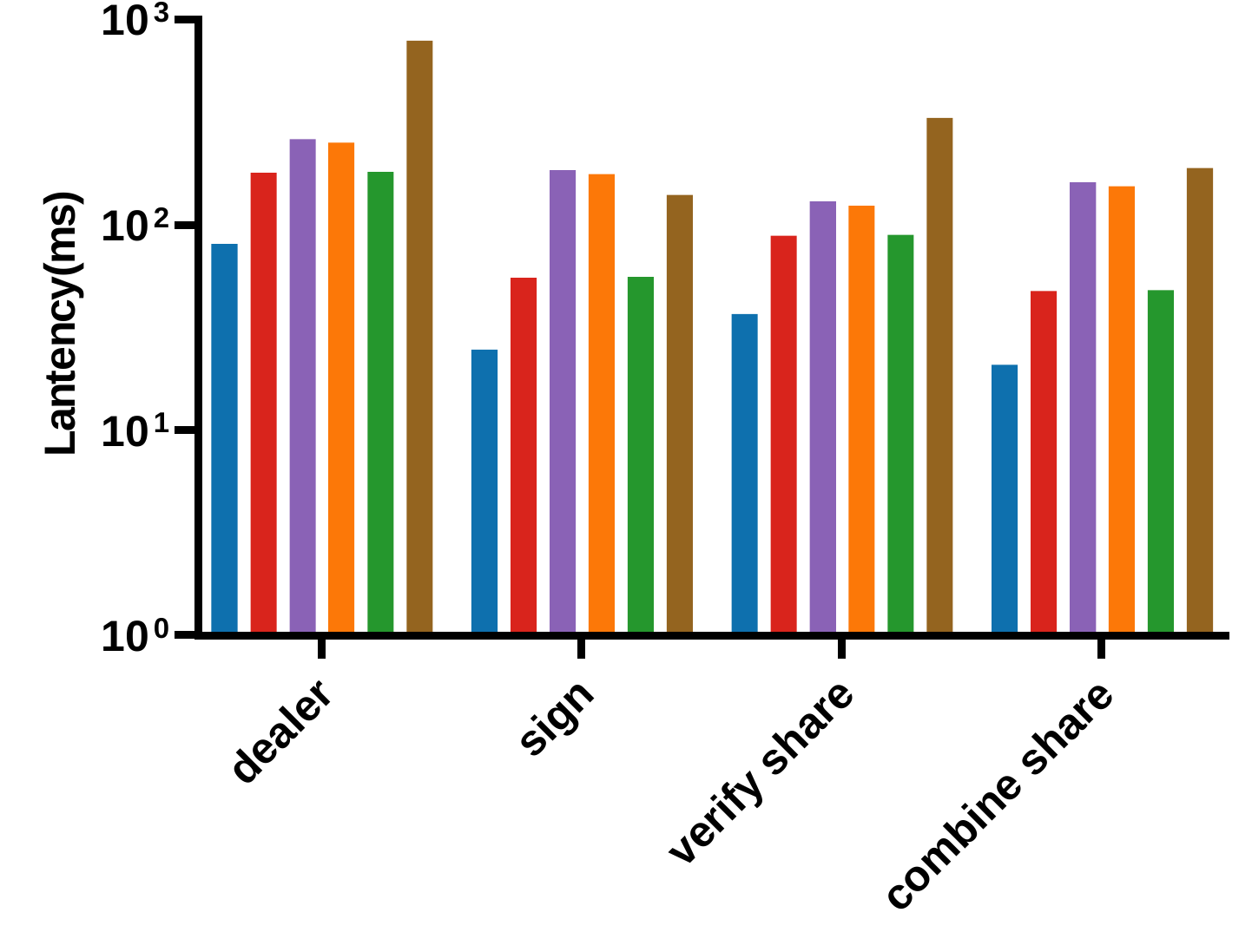}
    \label{fig:threshold_coin_flipping}
  }\\
  \subfloat[The size of public-key digital signature and threshold signature]{
    \includegraphics[width=0.47\linewidth]{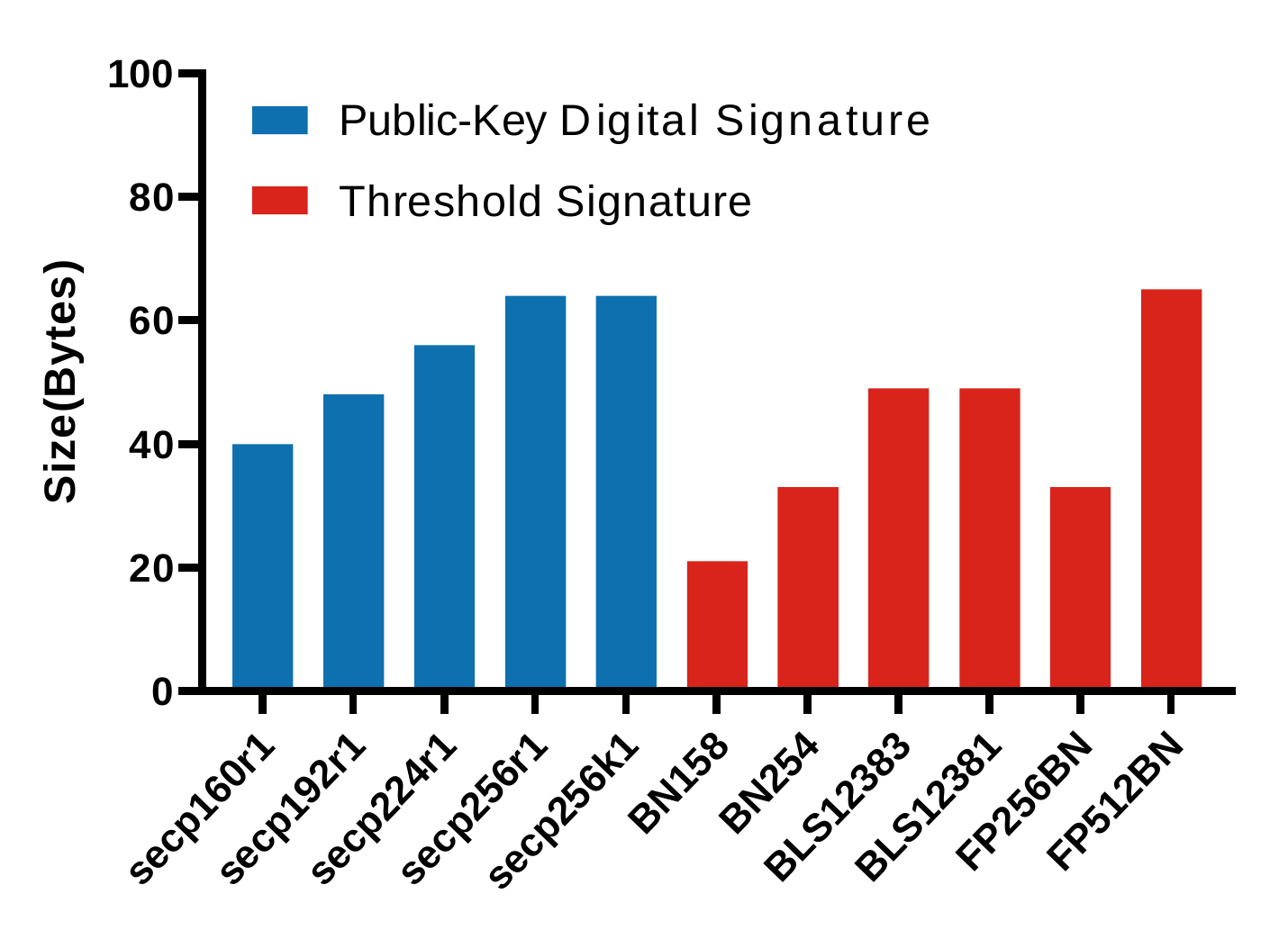}
    \label{fig:Size}
  }\hfill
  \subfloat[The latency and throughput of HoneyBadgerBFT with different cryptographic tools]{
    \includegraphics[width=0.47\linewidth]{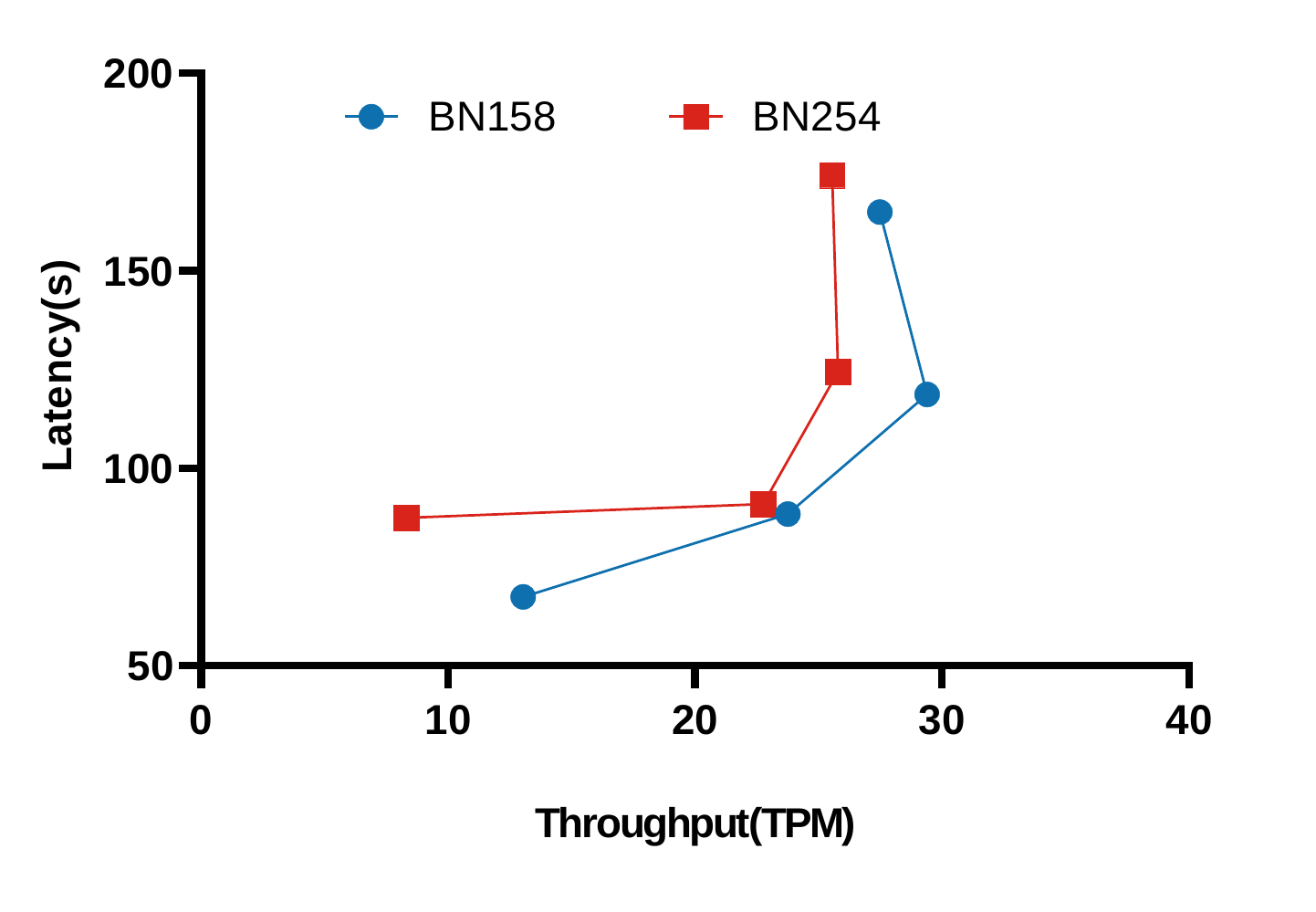}
    \label{fig:curves_consensus}
  }
  \caption{The latency and signature size of cryptographic tools and their impact on consensus}
  \label{fig:CryptoTools}
\end{figure}

For threshold cryptography, we test six representative deployment methods based on different curves provided by the MIRACL library\footnote{https://github.com/miracl}. Figures~\ref{fig:threshold_signature} and \ref{fig:threshold_coin_flipping} present the computation latency of basic operations in threshold signatures and threshold coin flipping. Among the six deployment methods, BN158 is the lightest, followed by BN254. Additionally, as shown in Figure~\ref{fig:Size}, we evaluate the signature sizes for the six deployment methods. BN158 produces the shortest threshold signature, measuring 21 bytes. For public-key digital signatures, we use five deployment methods based on different elliptic curves provided by the micro-ecc library\footnote{https://github.com/kmackay/micro-ecc}. We focus on the size of public-key digital signatures, as shown in Figure~\ref{fig:Size}. Secp160r1 generates the smallest digital signature, measuring 40 bytes. As illustrated in Figure~\ref{fig:curves_consensus}, using HoneyBadgerBFT as an example, we paired secp160r1 with BN158 and secp192r1 with BN254 to measure their impact on consensus latency and throughput. Results indicate that lighter curves enhance performance, with the BN158 group achieving a minimum latency reduction of 20 seconds and a maximum throughput increase of 4.7 transaction per minute (TPM). Based on these findings, we select secp160r1 and BN158 for subsequent experiments.

\subsection{Consensus Components}

\subsubsection{Broadcast Protocols}
\begin{figure}[h] 
\centering
  \subfloat[The latency of broadcast protocols vs. the number of parallel instances]{ \includegraphics[width=0.47\linewidth]{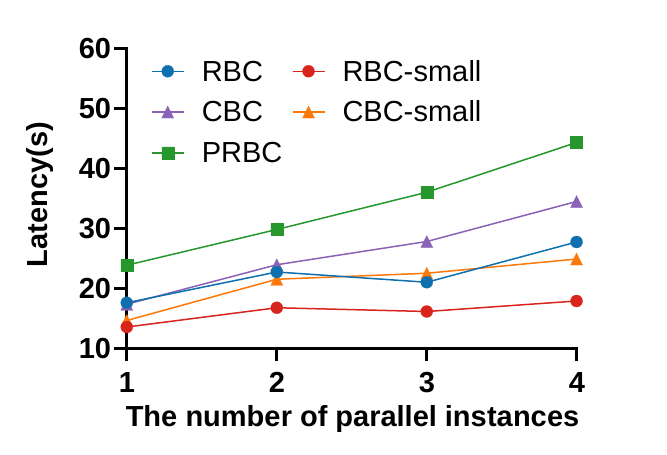}
\label{fig:BC-D}}\hfill
  \subfloat[The latency of broadcast protocols vs. proposal size]{  \includegraphics[width=0.47\linewidth]{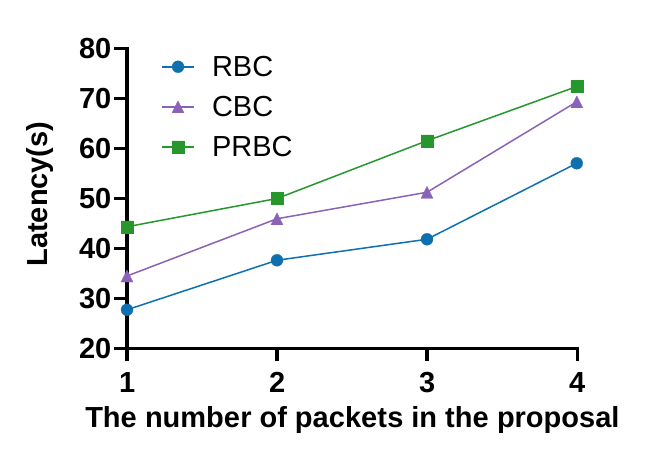}
\label{fig:BC-P}}\hfill
  \caption{The performance of broadcast protocols varying with parallel instances or proposal size}
  \label{fig:BroadcastProtocols} 
\end{figure}

For broadcast protocols, we evaluate RBC, PRBC, and CBC, along with their optimized versions for small broadcast values, RBC-small and CBC-small. As shown in Fig.~\ref{fig:BC-D}, regardless of the number of parallel instances, CBC and PRBC, which utilize threshold signatures, exhibit higher latency than RBC. RBC-small and CBC-small demonstrate greater stability across varying levels of parallelism compared to RBC and CBC. As parallelism increases, the advantages of them become more pronounced. When the parallelism is 4, RBC-small reduces latency by 9.8 seconds compared to RBC, improving performance by 35.5\%. Similarly, CBC-small reduces latency by 9.6 seconds compared to CBC, achieving a 27.8\% performance gain. As shown in Fig.~\ref{fig:BC-P}, we evaluate the latency of these three broadcast protocols with varying sizes of proposal. We represent the size of a proposal by the number of data packets it occupies. The latency ranking of the three broadcast protocols is consistent with that in Fig.~\ref{fig:BC-D}. As the size of the proposal increases, the performance gap between CBC and RBC grows larger, while the gap between CBC and PRBC becomes smaller. This indicates that the impact of cryptographic tools on performance is greater than that of message complexity.


\subsubsection{ABA}

\begin{figure}[h] 
\centering
  \subfloat[The latency of ABA vs. the number of parallel instances]{
    \includegraphics[width=0.47\linewidth]{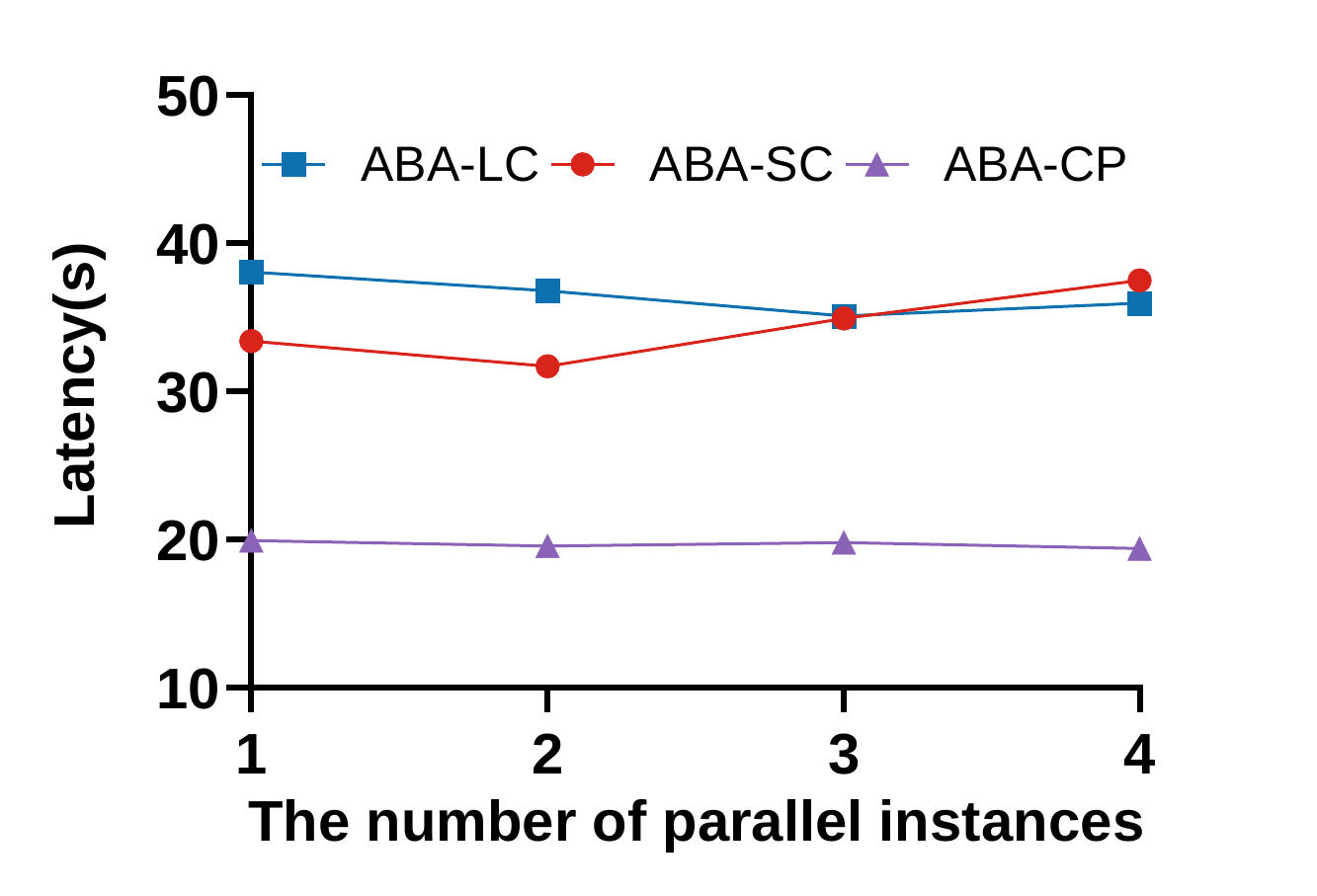}
    \label{fig:ABA-parallel}
  }\hfill
  \subfloat[The latency of ABA vs. the number of serial instances]{
    \includegraphics[width=0.47\linewidth]{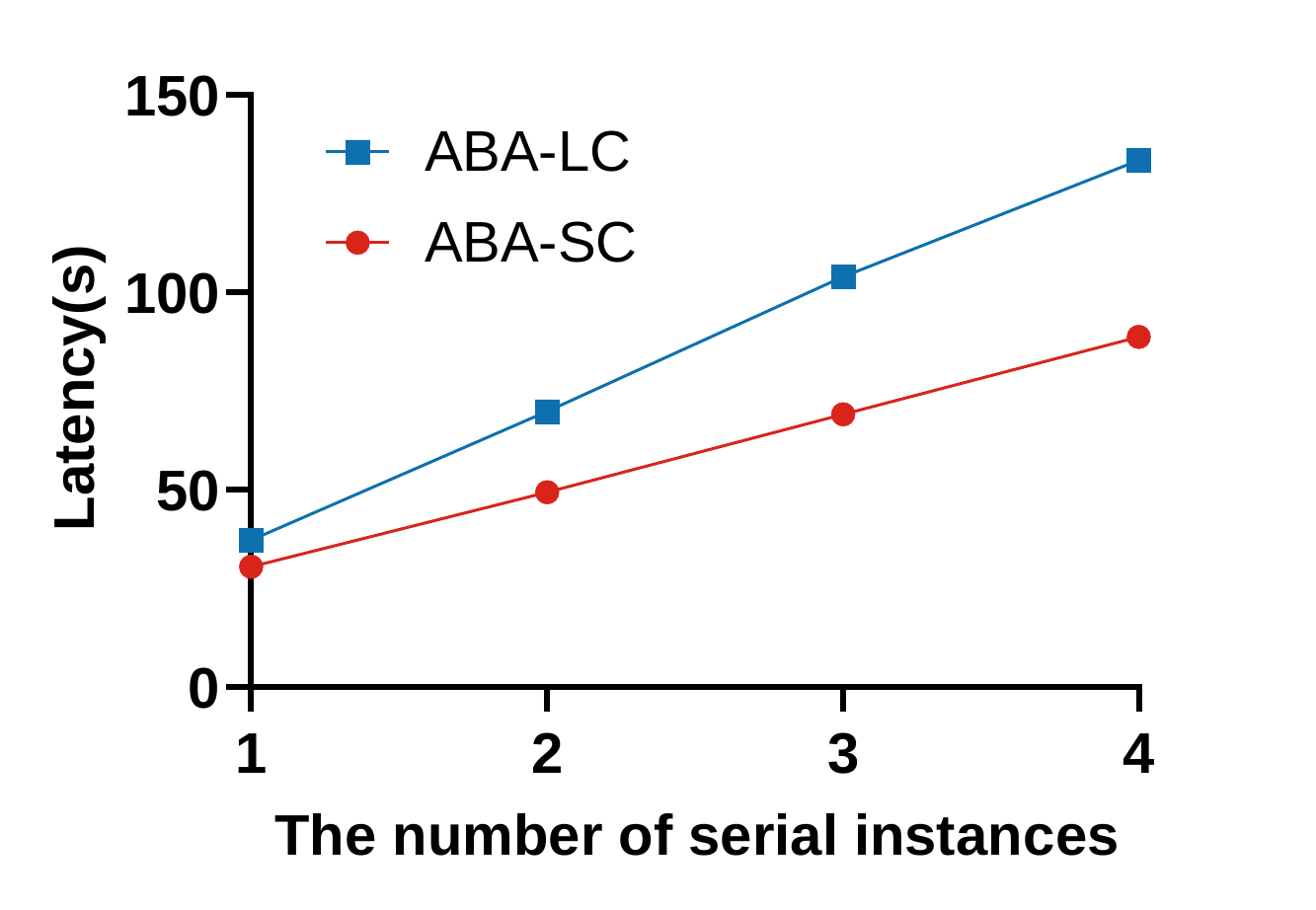}
    \label{fig:ABA-serial}
  }\hfill
  \caption{The performance of ABA varying with parallel instances or serial instances}
  \label{fig:ABA}
\end{figure}

We implement and compare three versions of ABA and focus on their latency associated with varying numbers of parallel or serial ABA instances. We use ABA-LC (local coin) to represent Bracha’s ABA, ABA-SC (shared coin) to represent Cachin’s ABA, and ABA-CP (coin flipping) to represent the ABA in BEAT. As shown in Fig.~\ref{fig:ABA-parallel}, we evaluate the impact of parallelism on latency. As the number of parallel instances increases, the latency gap between ABA-LC and ABA-SC becomes smaller, with ABA-LC outperforming ABA-SC at a parallelism level of 4, as ABA-SC requires more rounds to terminate. ABA-CP is more stable and exhibits lower latency compared to ABA-SC, as the computation delay required for threshold coin flipping is smaller than that for threshold signatures. As shown in Fig.\ref{fig:ABA-serial}, we evaluate the latency of ABA with varying numbers of serial instances. Running ABA serially corresponds to an ABA parallelism degree of 1. In this case, the latency of ABA-SC is lower than that of ABA-LC, consistent with the results observed in Fig.\ref{fig:ABA-parallel} for a parallelism degree of 1.

\subsection{Consensus}
We evaluate asynchronous BFT consensus in both single-hop and multi-hop network environments. The single-hop tests involve four nodes, while the multi-hop tests include 16 nodes distributed across four clusters. To avoid interference and facilitate both local and global consensus in the multi-hop setup, separate channels are used.
We evaluate five consensus protocols based on ConsensusBatcher: HoneyBadgerBFT-LC, HoneyBadgerBFT-SC, Dumbo-LC, Dumbo-SC, and BEAT. In our experiments, the shared coin versions consistently outperform their local coin counterparts. Therefore, we focus on the shared coin versions for both the ConsensusBatcher-based protocols and the baseline consensus protocols.
The three baseline consensus protocols, which do not employ ConsensusBatcher, are HoneyBadgerBFT-SC-baseline, Dumbo-SC-baseline, and BEAT-baseline.

\subsubsection{Single-Hop}
\begin{figure}[h] 
\centering
\includegraphics[width=0.94\linewidth]{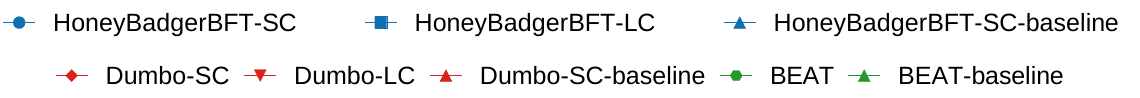}
    \label{fig:legend_consensus}
  \subfloat[The latency and throughput of 8 consensus protocols under single-hop scenarios]{
   \includegraphics[width=0.47\linewidth]{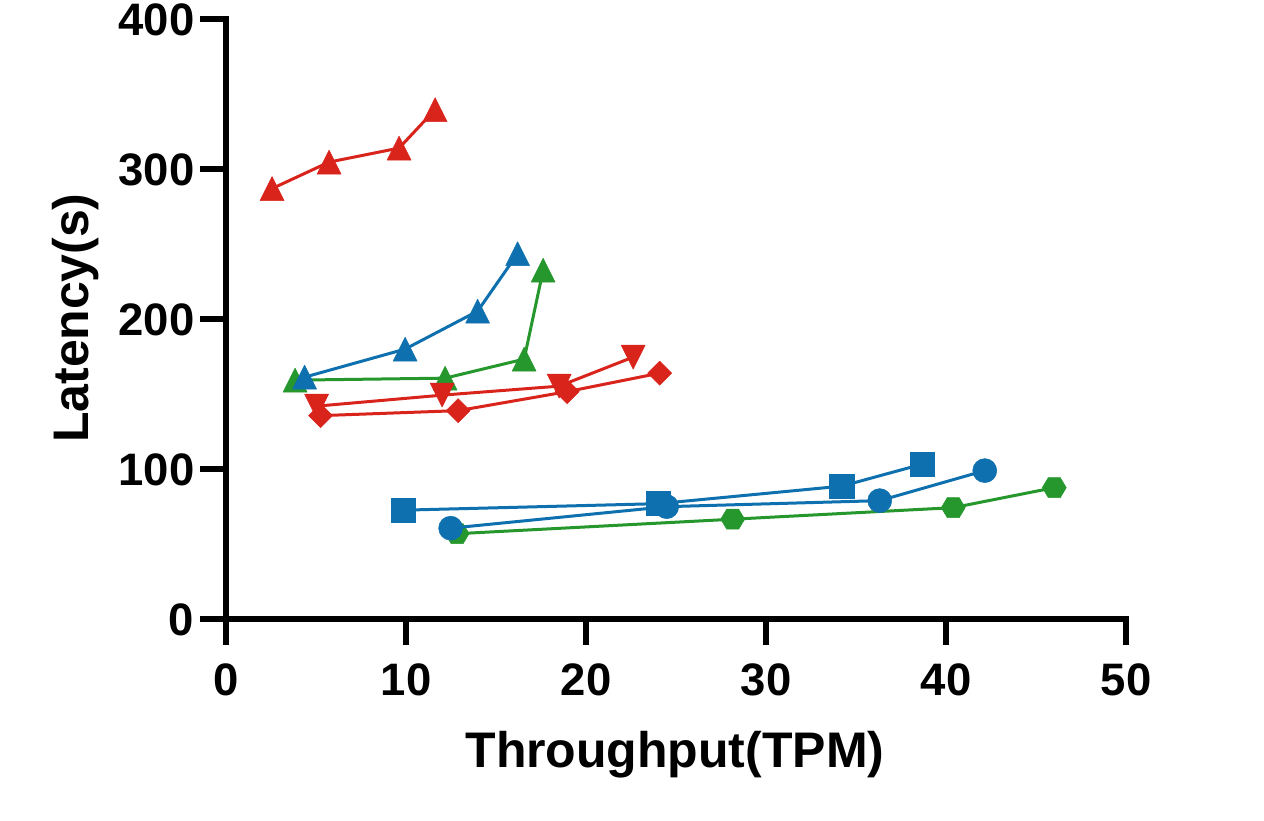}
\label{fig:single-hop}}\hfill
  \subfloat[The latency and throughput of 8 consensus protocols under multi-hop scenarios]{
    \includegraphics[width=0.47\linewidth]{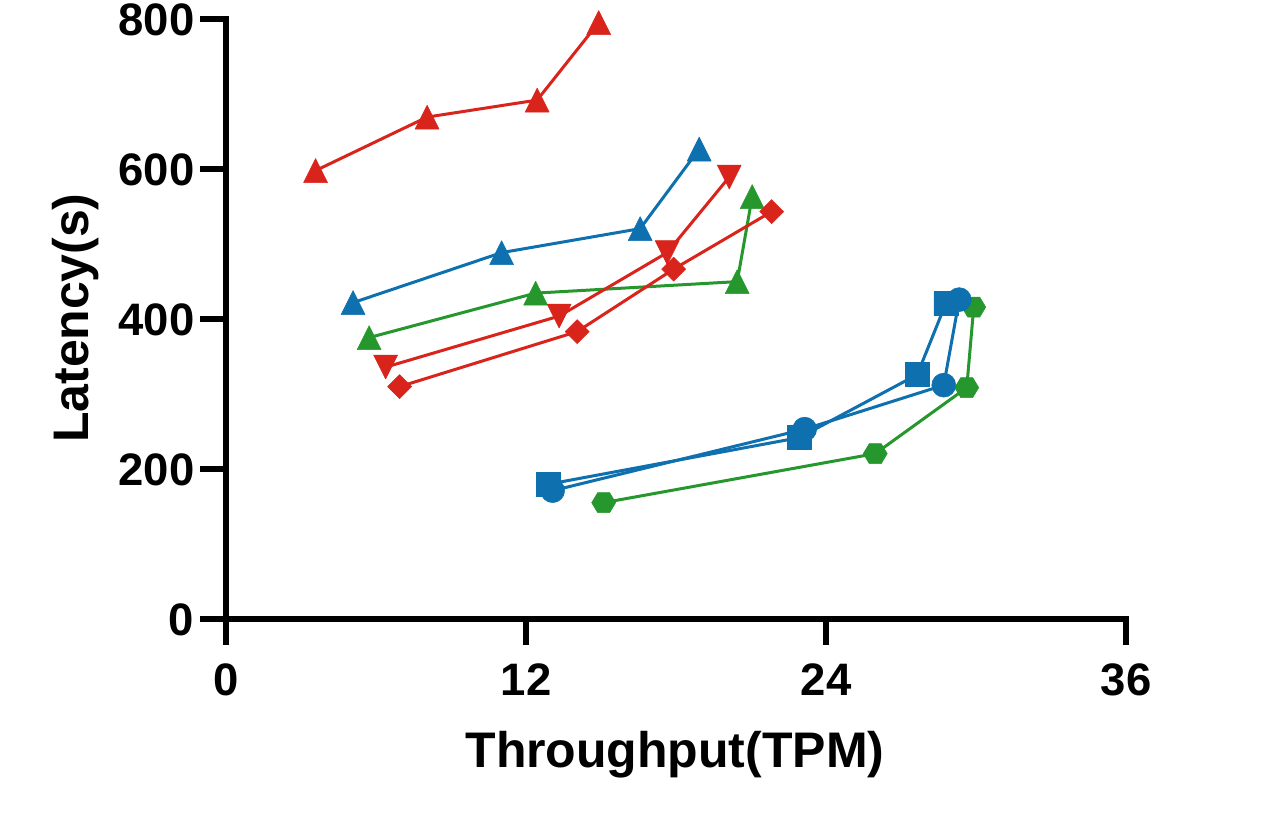}
\label{fig:multi-hop}}
\caption{The performance of 8 consensus protocols under single-hop or multi-hop scenarios}
  \label{fig:single-Consensus} 
\end{figure}

As shown in Fig.~\ref{fig:single-hop}, among the five consensus protocols based on ConsensusBatcher, BEAT achieves the best performance in both latency and throughput. Although Dumbo outperforms HoneyBadgerBFT in wired networks~\cite{guo2020dumbo}, HoneyBadgerBFT performs better in wireless environments due to Dumbo's more complex components. For both HoneyBadgerBFT and Dumbo, the local coin version of each consensus protocol performs slightly worse than the shared coin version. This is because message complexity has a greater impact than the threshold cryptographic tools, with one contributing factor being the key role played by our lightweight cryptographic tool implementation. The performance gap between the two versions of Dumbo is smaller than that between the two versions of HoneyBadgerBFT, indicating that ABA has less impact on Dumbo, leading to greater stability. Compared to baseline consensus protocols, the ConsensusBatcher-based consensus protocols show improvements in both latency and throughput. Specifically, latency is reduced by 52\% to 69\%, and throughput is increased by 50\% to 70\%. Additionally, the consensus protocols based on ConsensusBatcher are more stable than the baseline consensus protocols.

\subsubsection{Multi-Hop}

In multi-hop networks, BEAT maintains its lead in both latency and throughput, among the five consensus protocols based on ConsensusBatcher, as shown in Fig.~\ref{fig:multi-hop}. Dumbo continues to perform worse than HoneyBadgerBFT. Notably, Dumbo's throughput decreases in multi-hop networks compared to single-hop networks, whereas HoneyBadgerBFT's throughput increases due to its simpler process, which benefits from the multi-hop network's ability to improve throughput. The latency comparison between multi-hop and single-hop networks is not a straightforward doubling, as global consensus needs to wait for the majority of local consensus to complete, and proposals in global consensus involve proposals and signatures in local consensus. As in single-hop networks, the local coin version of each consensus protocol performs slightly worse than the shared coin version. However, the gap between the two versions becomes smaller, indicating that ABA's influence on consensus decreases in more complex network architectures. Compared to the baseline consensus protocols, the ConsensusBatcher-based consensus protocols show improvements in both latency and throughput. Specifically, latency is reduced by 48\% to 59\%, and throughput is increased by 48\% to 62\%. Performance improvements are less significant in multi-hop networks because of their complexity.

\section{Conclusion}
We propose ConsensusBatcher, a protocol for batching parallel consensus components, adapting asynchronous BFT consensus (e.g., HoneyBadgerBFT, BEAT, Dumbo) from wired to wireless networks. An open-source testbed is developed to evaluate performance.
Future work focuses on optimizing asynchronous BFT consensus through improved communication protocols and application-specific integration (e.g., satellite networks). We will also explore combining embodied AI with BFT consensus for enhanced AI system security.

\bibliographystyle{IEEEtran}
\bibliography{main}

\end{document}